\documentclass[noshowframe]{imsart}

\RequirePackage{amsthm,amsmath,amsfonts,amssymb}
\RequirePackage[authoryear]{natbib}%% uncomment this for author-year citations
\RequirePackage[colorlinks,citecolor=blue,urlcolor=blue]{hyperref}
\RequirePackage{graphicx}

\arxiv{2506.16175}
\startlocaldefs
\theoremstyle{plain}
\newtheorem{axiom}{Axiom}
\newtheorem{claim}[axiom]{Claim}
\newtheorem{theorem}{Theorem}[section]
\newtheorem{lemma}[theorem]{Lemma}

\usepackage{comment}
\usepackage{graphicx} % Required for inserting images
\usepackage{amsmath}
\usepackage{multicol}
\usepackage{makecell}
\usepackage{amssymb}
\usepackage{amsfonts,amsthm,lineno,placeins,xcolor}
\usepackage{comment}
\usepackage{standalone}
\usepackage{lscape}
\DeclareMathOperator{\EX}{\mathbb{E}}% expected value
\DeclareMathOperator{\Cov}{Cov}% expected value
\newcommand{\indep}{\perp \!\!\! \perp}
\theoremstyle{definition}
\newtheorem{definition}[theorem]{Definition}
\newtheorem*{example}{Example}
\newtheorem*{fact}{Fact}
\newcommand{\revision}[1]{\textcolor{black}{#1}}
\theoremstyle{remark}
\newtheorem{case}{Case}
\endlocaldefs

\begin{document}
\begin{frontmatter}
\title{Independent vector analysis -- an introduction for statisticians}
%\title{A sample article title with some additional note\thanksref{t1}}
\runtitle{Independent vector analysis - an introduction for statisticians}
%\thankstext{T1}{A sample additional note to the title.}

\begin{aug}
%%%%%%%%%%%%%%%%%%%%%%%%%%%%%%%%%%%%%%%%%%%%%%%
%% Only one address is permitted per author. %%
%% Only division, organization and e-mail is %%
%% included in the address.                  %%
%% Additional information can be included in %%
\author[A]{\fnms{Miro}~\snm{Arvila}\ead[label=e1]{miro.t.arvila@jyu.fi}\orcid{0009-0001-8737-0754}},
\author[B]{\fnms{Klaus}~\snm{Nordhausen}\ead[label=e2]{klaus.nordhausen@helsinki.fi}\orcid{0000-0002-3758-8501}}
\author[A]{\fnms{Mika}~\snm{Sipil\"a}\ead[label=e3]{mika.e.sipila@jyu.fi}\orcid{0000-0002-5912-840X}}
\and
\author[A]{\fnms{Sara}~\snm{Taskinen}\ead[label=e4]{sara.l.taskinen@jyu.fi}\orcid{0000-0001-9470-7258}}
%%%%%%%%%%%%%%%%%%%%%%%%%%%%%%%%%%%%%%%%%%%%%%
%% Addresses                                %%
%%%%%%%%%%%%%%%%%%%%%%%%%%%%%%%%%%%%%%%%%%%%%%
\address[A]{Department of Mathematics and Statistics, University of Jyv\"askyl\"a, Finland\printead[presep={,\ }]{e1,e3,e4}}

\address[B]{Department of Mathematics and Statistics, University of Helsinki, Finland\printead[presep={,\ }]{e2}}
\runauthor{M. Arvila et al.}
\end{aug}

\begin{abstract}
Blind source separation (BSS), particularly independent component analysis (ICA), has been widely used in various fields of science such as biomedical signal processing to recover latent source signals from the observed mixture. While ICA is typically applied to individual datasets, many real-world applications share underlying sources across datasets. Independent vector analysis (IVA) extends ICA to jointly analyze multiple datasets by exploiting statistical dependencies across them. While various IVA methods have been presented in signal processing literature, the statistical properties of methods remains largely unexplored. This article introduces the IVA model, numerous density models used in IVA, and various classical IVA methods to statistics community highlighting the need for further theoretical developments.

\end{abstract}


%\begin{keyword}[class=MSC]
%\kwd[Primary: 62-02]
%\kwd[Secondary:]62EXX Statistical distribution theory?
%\kwd{62Hxx Multivariate analysis? (05,10)}
%\kwd{62H99 None of the above, but in this section?}
%\end{keyword}


\end{frontmatter}
%%%%%%%%%%%%%%%%%%%%%%%%%%%%%%%%%%%%%%%%%%%%%%
%% Please use \tableofcontents for articles %%
%% with 50 pages and more                   %%
%%%%%%%%%%%%%%%%%%%%%%%%%%%%%%%%%%%%%%%%%%%%%%
%\tableofcontents

\begin{comment}
\textcolor{red}{Do we need notations as most are not anymore used and those that are used are defined in the paper?}
\section*{Notations}
List of used notations in the paper \\
$\EX(\cdot)$ expected value operator\\
$\odot$ Hadamard product\\
$\otimes$ Kronecker product\\
$\oplus$ direct sum \\
$\Gamma(\cdot)$ Gamma function  \\
$^\top$ transpose \\
$\indep$ independency of random variables \\
$\mathcal{I}(\cdot)$ Mutual information \\
$\mathcal{H}(\cdot)$ Differential entropy \\
$\mathcal{H}(\cdot:\cdot,\cdot)$ Differential cross entropy\\
$D_{KL}(\cdot|\cdot)$ Kullback-Leibler divergence\\
$\mathcal{J}(\cdot)$ Negentropy\\
$\mathbb{N}$ Nonnegative natural numbers\\
$\mathbb{R}$ real numbers \\
$\delta_{i,j}$ Kronecker delta for indeces $i$ and $j$
\end{comment}

\section{Introduction}\label{sec:intro}
Blind source separation (BSS) has been studied for several decades across various fields, including computer science, signal processing, and statistics. The goal of BSS is to separate a set of source signals from observed mixtures, using little or no prior information about the sources or the mixing process. One of the most well-known BSS methods is independent component analysis \citep[ICA,][]{Comon1994}, which has been extensively studied in the literature.

The statistical properties of various ICA estimators are now well understood, and ICA has found applications in numerous domains such as astrophysics, telecommunications, medical signal processing, and finance, to name a few.

In biomedical signal processing, independent component analysis (ICA) is typically applied to data from a single subject. However, in many studies, the same experimental protocol is carried out across multiple subjects. For example, in an electroencephalogram (EEG) study, all subjects are asked to perform the same tasks in a controlled environment, with the goal of understanding how different regions of the brain (represented by latent sources) are activated. 

In EEG, the observed signals are the result of electrodes placed on the scalp, which simultaneously record the electrical activity emitted by all regions of the brain. Although ICA can be applied separately to each subject, this approach ignores the fact that we can reasonably assume that brain function is similar across individuals, and thus the underlying independent sources should be comparable. This shared structure across subjects provides additional information that could be leveraged. However, due to anatomical differences between individuals and possible variations in electrode placement, the mixing of the signals is likely to differ from one subject to another.

To address such scenarios, independent vector analysis \citep[IVA,][]{Taesu2006IVA} was introduced as a generalization of ICA to multiple datasets, by exploiting statistical dependencies across them. IVA has been actively studied in the signal processing community for nearly two decades but remains relatively unknown in the statistical literature. It has already been successfully applied in a range of fields, including audio and speech signal separation \citep{Lee2007speech,Brendel2020spat,Amor2021}, medical signal processing, for example, in EEG and \revision{functional magnetic resonance imaging} (fMRI) studies \citep{LEE2008fmri,Adali2015,Adali2018}, and in wireless communication \citep{Luo2022com,Luo2022mimo,Li2022comIVA}.

The purpose of this article is to introduce IVA to the statistical community. We believe that IVA is a valuable and relevant area where contributions from statisticians could be highly beneficial, especially since many statistical properties of the methods remain unexplored.

The paper is organized as follows. Section~\ref{sec:ICA} provides a brief overview of independent component analysis (ICA), beginning with the model formulation and followed by various approaches to solve the ICA problem. A comprehensive review of IVA is presented in Section~\ref{sec:IVA}. Section~\ref{sec:related} discusses methods that are closely related to IVA and highlights their differences. In Section~\ref{sec:application}, we illustrate the use of IVA through a real-life application involving mixed color images \revision{and fMRI data}. The paper concludes with a discussion \revision{and some open problems} in Section~\ref{sec:discussion}.

\section{Independent component analysis}\label{sec:ICA}

\subsection{Independent component model and unmixing matrix functionals}

Before introducing the independent vector model, let us recall the independent component (IC) model and some common methods for ICA. The basic IC model assumes that the $p$-variate observed components of $\boldsymbol{x}_i = (x_{i1},\dots,x_{ip})^\top$ are linear combinations of latent independent components of $\boldsymbol{s}_i = (s_{i1},\dots,s_{ip})^\top$.
Hence, the model can be written as

%The goal of ICA is to find a linear transformation for the observed signals so that the estimated components are statistically as independent as possible \citep{COMON1994287}. Assume now that each $p$-variate observed signal $\boldsymbol{x}_1,\dots,\boldsymbol{x}_n$ follow the IC-model 

%\textcolor{olive}{Should we still say somewhere that $\mathbf{x}_i = (x_{i1}, \dots, x_{ip})^\top$ and $\mathbf{s}_i = (s_{i1}, \dots, s_{ip})^\top$ for clarification when introducing $\mathbf{x}_i$ and $\mathbf{s}_i$?}
\begin{align}\label{eq:icmodel}
\boldsymbol{x}_i = \boldsymbol{\Omega}\boldsymbol{s}_i + \boldsymbol{\mu}, \; \; \text{$i = 1,\dots,n$},
\end{align}
where $\boldsymbol{\mu}$ is a $p$-variate location parameter, $\boldsymbol{\Omega}$ is a full rank $p \times p$ mixing matrix, and $\boldsymbol{s}_1,\dots,\boldsymbol{s}_n$ are independent and identically distributed $p$-variate random vectors. The structure of the ICA model can be seen in Figure \ref{ICAMODELpic} \revision{in \ref{ICA figure}}.

In ICA one usually assumes that
\begin{itemize}
\item [(A1)] The components $s_{i1},\dots,s_{ip}$ are independent.
\item [(A2)] $\EX(\boldsymbol{s}_i) = \boldsymbol{0}$ and $\EX(\boldsymbol{s}_i\boldsymbol{s}_i^\top) = \boldsymbol{I}_p$, where $\boldsymbol{I}_p$ is a $p\times p$ identity matrix.
\item [(A3)] At most one of the components $s_{i1},\dots,s_{ip}$ of $\boldsymbol{s}_i$ has a normal distribution.
\end{itemize}

The first assumption (A1) is natural, as the goal is to estimate independent components. Assumption (A2) requires the existence of second moments, thereby fixing the location and scale of the components. Assumption (A3) is essential for identifiability. For example, if $\boldsymbol{s}_i \sim \mathcal{N}_2(\boldsymbol{0}, \boldsymbol{I}_2)$, then $\boldsymbol{U}\boldsymbol{s}_i \sim \mathcal{N}_2(\boldsymbol{0}, \boldsymbol{I}_2)$ for any orthogonal $2 \times 2$ matrix $\boldsymbol{U}$, implying that the independent components are not uniquely defined.

Even under assumptions (A1)–(A3), the independent component (IC) model \eqref{eq:icmodel} is not uniquely determined. Let $\boldsymbol{J}$ be a $p \times p$ sign-change matrix (a diagonal matrix with $\pm 1$ on the diagonal), and let $\boldsymbol{P}$ be a $p \times p$ permutation matrix (each row and column contains a single entry equal to one, with all other entries zero). Then the IC model can be rewritten as
\[
\boldsymbol{x}_i = \boldsymbol{\Omega}\boldsymbol{s}_i + \boldsymbol{\mu} 
= \boldsymbol{\Omega} \boldsymbol{J}\boldsymbol{P} (\boldsymbol{P}^\top \boldsymbol{J} \boldsymbol{s}_i) + \boldsymbol{\mu} 
= \boldsymbol{\Omega}^\ast \boldsymbol{s}_i^\ast + \boldsymbol{\mu},
\]
where $\boldsymbol{\Omega}^\ast = \boldsymbol{\Omega} \boldsymbol{J}\boldsymbol{P}$ and $\boldsymbol{s}_i^\ast = \boldsymbol{P}^\top \boldsymbol{J} \boldsymbol{s}_i$. This means that the components can only be recovered up to sign and permutation. However, in most ICA applications, this level of identifiability is sufficient and not considered a limitation.

The main idea of ICA is to recover the latent independent components $\boldsymbol{s}_i$ and the mixing matrix $\boldsymbol{\Omega}$ only using the observed signals $\boldsymbol{x}_i$, $i = 1,\dots,n$. By abuse of notation this can be done by finding a $p \times p$ unmixing matrix $\boldsymbol{W}$ such that 
\[
\boldsymbol{s}_i = \boldsymbol{W}(\boldsymbol{x}_i-\boldsymbol{\mu}), \; \; \text{$i = 1,\dots,n$}.
\]
If $\boldsymbol x_1,\dots,\boldsymbol x_n$ is a random sample from a cumulative distribution function $F_{\boldsymbol x}$, then the population unmixing matrix, which we wish
to estimate, is defined as the value of an unmixing matrix functional at $F_{\boldsymbol x}$, that is, 
$\boldsymbol{W}(\boldsymbol x)$, where $\boldsymbol{x}$ is a random variable from distribution with cdf of $F_{\boldsymbol{x}}$. We say that the $p \times p$ matrix-valued functional $\boldsymbol{W}(\boldsymbol{x})$ is an unmixing matrix functional if (i) $\boldsymbol{W}(\boldsymbol{x})\boldsymbol{x}$ has independent components in the IC-model \eqref{eq:icmodel} and (ii) $\boldsymbol{W}(\boldsymbol{x})$ is affine equivariant, i.e., $\boldsymbol{W}(\boldsymbol{Ax+b}) =\boldsymbol{PJW}(\boldsymbol{x})\boldsymbol{A}^{-1}$, where $\boldsymbol{A}$ is any invertible $p \times p$ matrix and $\boldsymbol{b}$ is any $p$-vector \citep{miettinen2015}. The $p \times p$ matrix $\boldsymbol{P}$ is a permutation matrix and $\boldsymbol{J}$ is a sign-change matrix as defined above.

In many classical ICA methods, the first step involves preprocessing the data through whitening. Whitening transforms the data so that the variables are uncorrelated, have unit variance, and are centered at the origin. The main advantage of this step is that, after whitening, it is sufficient to find an orthogonal matrix to recover the independent components.

More precisely, if $\boldsymbol{x}$ follows the IC model, then its mean vector is $\boldsymbol{\mu}$ and its covariance matrix is $\Cov(\boldsymbol{x}) = \boldsymbol{\Omega\Omega}^{\top}$. The whitened vector is defined as
\[
\boldsymbol{x}_{{st}} = \Cov(\boldsymbol{x})^{-1/2}(\boldsymbol{x} - \boldsymbol{\mu}),
\]
and one can show that independent components can be recovered by applying an orthogonal rotation to $\boldsymbol{x}_{{st}}$, that is,
\[
\boldsymbol{s} = \boldsymbol{U}\boldsymbol{x}_{{st}},
\]
where $\boldsymbol{U}$ is a $p \times p$ orthogonal matrix \citep{miettinen2015}. The unmixing matrix $\boldsymbol{W}$ is then given by $\boldsymbol{W} = \boldsymbol{U} \Cov(\boldsymbol{x})^{-1/2}$. The different ICA methods differ mainly in the way the orthogonal matrix $\boldsymbol{U}$ is estimated. %The following methods will be written using $\boldsymbol{W}$ instead of $\boldsymbol{U}$, because it is not known if there is same kind of result for IVA as ICA above.
The following methods will be presented using $\boldsymbol{W}$ instead of $\boldsymbol{U}$, as it is not known whether a similar result holds for IVA as for ICA above.

In BSS, the location parameter $\boldsymbol{\mu}$ is generally considered a nuisance parameter. Without loss of generality, we therefore assume in the following that $\EX(\boldsymbol{x}) =\boldsymbol{\mu} = \boldsymbol{0}$.

\revision{As a side note ICA has been studied extensively with different data types. For recent studies see \citet{virta2017independent,VirtaLiNordhausenOja2020}, where the authors extended ICA to tensor and functional data.}

\subsection{Entropy-based ICA methods}  

The ICA problem can be addressed using a variety of approaches. For reviews, see \citet{HyvarinenICAbook,ComonJutten2010,NordhausenOja2018}. These approaches are typically divided into two main categories.

The first category comprises algebraic methods, such as fourth order blind identification \citep[FOBI,][]{Cardoso1989} and joint approximate diagonalization of eigen-matrices \citep[JADE,][]{CardosoSouloumiac1993}. In this paper, we do not focus on algebraic methods. For further references on this class of methods, see \citet{OjaSirkiaEriksson2006, miettinen2015, nordhausen2019overview, NordhausenRuizGazen2022}.

The second category includes methods that aim to minimize or maximize the cost function. Examples include maximum likelihood estimation (MLE) \citep{Pearlmutter1996, Cardoso1997, karvanen2000maximum}, mutual information minimization \citep{Calhoun2002}, and negentropy maximization \citep{Bingham2000}. These optimization-based methods are briefly reviewed below, starting with MLE. The density for the observed signal $\boldsymbol{x}$ from the IC model is now   
\begin{align*}
    p_{\boldsymbol{x}}(\boldsymbol{x}) = |\det \boldsymbol{W}| p_{\boldsymbol{s}}(\boldsymbol{s}) = |\det \boldsymbol{W}| \prod_{j=1}^p p_{s_j}(s_j) = |\det \boldsymbol{W}| \prod_{j=1}^p p_{s_j}(\boldsymbol{w}_j^\top \boldsymbol{x}),
\end{align*}
where $\boldsymbol{W} = (\boldsymbol{w}_1, \dots , \boldsymbol{w}_p)^\top =  \boldsymbol{\Omega}^{-1}$ and $p_{s_j}$ are the densities of the independent components. 
The cost function used in maximum likelihood is the log-likelihood function which is
\begin{align}\label{eq:ICAml}
    \log p_{\boldsymbol{x}}(\boldsymbol{x}) = \sum_{j=1}^p \log p_{s_j}(\boldsymbol{w}_j^\top \boldsymbol{x}) 
    + |\det \boldsymbol{W}|.
\end{align}
The problem with MLE is that the densities $p_{s_j}$ have to be known in order to successfully estimate $\boldsymbol{W}$. In practice, the densities are usually not known, and if wrong assumptions on them are made, then also the estimate of $\boldsymbol{W}$ may be incorrect. Fortunately, there are methods that suffer less from unknown source densities. One approach is to assume some parametric form of source densities, for which the parameters can be estimated using MLE. This approach is used for example in Pearson-ICA \citep{karvanen2000maximum}, ProDensICA \citep{Hastie2002ProDensICA} and logconcave density ICA \citep{Samworth2012logconcICA}. Other commonly used approaches include mutual information and negentropy based methods, which are closely related to MLE and reviewed in the following.

Mutual information is another way to approach the ICA problem. In ICA context, mutual information is used as a metric which measures the distance between a distribution and the product of its marginal distributions. One can choose the first distribution to be the joint probability distribution of the source estimates $p_{\boldsymbol{s}}(\hat{s}_1,\dots,\hat{s}_p)$ and the second distribution to be the product of the marginal probability distributions of $\prod_{j=1}^p p_{s_j}(\hat{s}_j)$, where the true source distributions $p_{\boldsymbol{s}}$ and $p_{s_j}$ are unknown. Here $\hat{s}_j = \boldsymbol{w}_j^\top \boldsymbol{x}$, $j=1,\dots,p$ are the estimated source components. If the Kullback-Leibler divergence is zero for the joint distribution and the product of the marginal distributions, then the distributions are the same, which implies the independence between the source estimates. More information about these measures can be found in the Appendix \ref{information theory}.

The ICA cost function can be derived using mutual information as follows. Below the index $i$ is dropped, because the source estimates are considered as population-level variables. The cost function is
\begin{align}
    \mathcal{I}_{ICA} &= D_{KL}(p_{\boldsymbol{s}}(\hat{s}_1,\dots,\hat{s}_p)|\prod_{j=1}^p p_{s_j}(\hat{s}_j))\nonumber \\
    &= -\int p_{\boldsymbol{s}}(\hat{s}_1,\dots,\hat{s}_p)\log(\prod_{j=1}^p p_{s_j}(\hat{s}_j))d\boldsymbol{s} - \mathcal{H}(\hat{\boldsymbol{s}})\nonumber \\
    &= \sum_{j=1}^p\left(-\int p_{\boldsymbol{s}}(\hat{s}_1,\dots,\hat{s}_p)\log(p_{s_j}(\hat{s}_j))d\boldsymbol{s}\right) - \mathcal{H}(\boldsymbol{Wx})\nonumber \\
    &= \sum_{j=1}^p \mathcal{H}(\hat{s}_j) - \log|\det \boldsymbol{W}| - \mathcal{H}(\boldsymbol{x}) = \sum_{j=1}^p \mathcal{H}(\boldsymbol{w}_j^{\top}\boldsymbol{x}) - \log|\det \boldsymbol{W}| - \mathcal{H}(\boldsymbol{x}) \label{ICAcost},
\end{align}
where $\mathcal{H}(\cdot)$ denotes entropy.
As the distributions $p_{s_j}$ are unknown one has to choose prior source models for them. The cost function is then minimized with respect to the unmixing matrix $\boldsymbol{W}$. Note that $\mathcal{H}(\boldsymbol{x})$ is a constant with respect to the unmixing matrix $\boldsymbol{W}$ and can be dropped.

Mutual information and maximum likelihood are closely connected.
One can show that mutual information is equal to the negative log-likelihood if the source probability distributions were the true distribution~\citep{HYVARINEN2000411}. This can be seen by taking the expected value of negative (\ref{eq:ICAml}), which gives
\begin{align*}
    \EX(-\log p_{\boldsymbol{x}}(\boldsymbol{x}))=-\sum_{j=1}^p\EX(\log p_{s_j}(\boldsymbol{w}_j^\top\boldsymbol{x}))-|\det \boldsymbol{W}|.
\end{align*}
Now, if $p_{s_j}$ is equal to the actual probability density of $\boldsymbol{w}_j^\top\boldsymbol{x}$ then the first term is equal to $\mathcal{H}(\boldsymbol{w}_j^{\top}\boldsymbol{x})$. Thus, the likelihood is equal to mutual information up to constant and sign change.

Another way to approach the ICA problem is to use cost functions that measure non-Gaussianity. 
One popular choice for this is negentropy. The cost function for negentropy is
\begin{align*}
    \mathcal{J}(\hat{\boldsymbol{s}}) = \mathcal{H}(\hat{\boldsymbol{s}}_{gauss}) - \mathcal{H}(\hat{\boldsymbol{s}}),
\end{align*}
where $\boldsymbol{\hat{s}}_{gauss}$ is a Gaussian random vector of the same covariance matrix as $\hat{\boldsymbol{s}}$. More about negentropy can be found in the Appendix \ref{negen}.

One problem when using negentropy in optimization is that it is computationally heavy to calculate. Thus, different approximations for negentropy have been developed to make computations easier \citep{Jones1987,Hyvarinen1997negappox}. Two of such approximations are explained in the Appendix \ref{negen} with more details. Other approximations also have been used in literature, for example, \citet{Prasad2005} proposed an approximation using generalized higher-order statistics of different non-quadratic, non-linear functions and \citet{Chang01062008} approximated negentropy based on the maximum entropy principle to measure non-Gaussianity of data sequences.

To maximize negentropy, one can use, for example, a simple gradient algorithm based on a negentropy approximation. Another popular approach involves fixed-point algorithms. One of the most widely used fixed-point methods is the FastICA algorithm \citep{HyvarinenOja:1997}. For a detailed discussion on nonlinearities, negentropy approximations, practical choices of nonlinearities, and properties of the fixed-point algorithm, see \citet{Hyvarinen:1999}. \revision{For further details about different variants of FastICA algorithm see \citet{Nordhausenetal:2011,MiettinenNordhausenOjaTaskinen2014, miettinen2015,Miettinenetal:2017}.} 
In Section \ref{sec:IVA}, where the IVA model is discussed, we will use the same term nonlinearity. It should be noted that in some IVA literature, the term contrast function is used interchangeably with the nonlinearity function commonly used in ICA. Additionally, in some IVA papers, the term contrast function is used to refer to the cost function that is minimized. For these reasons, we will use the term nonlinearity throughout this paper.

Just like in maximizing negentropy, the same algorithms can be used for MLE estimations or mutual information. This follows from the fact that mutual information can be expressed using negentropy. This is stated in \ref{negen}. Another possibility is to use Infomax principle which was derived from a neural network point of view in \cite{Bell1995,Nadal1995}. Infomax principle works because network entropy maximization, i.e. Infomax, is equivalent to the maximum likelihood estimation \citep{Pearlmutter1996,Cardoso1997}.

\section{Independent vector analysis}\label{sec:IVA}

\begin{table}[htb]
\centering
\caption{Table of Notations}
\begin{tabular}{ll}
\hline
\textbf{Symbol} & \textbf{Meaning} \\
\hline
$(\cdot)^\top$ & Transpose \\
$\EX(\cdot)$ & Expected value\\
$D_{KL}(\cdot|\cdot)$ & Kullback-Leibler divergence\\
$\mathcal{H}(\cdot)$ & Entropy \\
$\mathcal{J}(\cdot)$ & Negentropy \\
$\phi(\cdot)$ & Multivariate score function\\
$\mathbf{H}$ & Hessian matrix\\
$\Gamma(\cdot)$ & Gamma function \\
$\delta_{l,a}$ & Kronecker delta i.e $\delta_{l,a} = 0$ if $l \neq a$ otherwise $\delta_{l,a} = 1$\\
$\boldsymbol{Y}_{gauss}$ & Gaussian random variable with same mean and covariance as $\boldsymbol{Y}$\\
$ Y \indep Z$ & $Y$ and $Z$ are independent\\
$n$ & Sample size for each dataset\\
$K$ & The number of datasets \\
$\boldsymbol{\mu}^{[k]}$ & Location of dataset $k$, usually assumed zero\\
$\boldsymbol{s}_i^{[k]} = (s_{i1}^{[k]},\dots,s_{ip}^{[k]})^\top$ & The $i$th sample of latent independent component from dataset $k$\\
$s_{ij}^{[k]}$ & $j$th component of $i$th sample from $k$th dataset\\
$\boldsymbol{\Omega}^{[k]}$ & The mixing matrix of dataset $k$\\
$\boldsymbol{x}_i^{[k]} = (x_{i1}^{[k]},\dots,x_{ip}^{[k]})^\top$ & The observed $i$th sample of dataset $k$\\
$x_{ij}^{[k]}$ & $j$th component of $i$th sample from $k$th dataset\\
$\boldsymbol{W}^{[k]}$ & The unmixing matrix of dataset $k$\\
$\boldsymbol{\hat{s}}_j = (\hat{s}_j^{[1]},\dots,\hat{s}_j^{[K]})^\top$ & The $j$th estimated source component vector\\
$\text{vec}_{\boldsymbol{\alpha}}(\boldsymbol{y})$ & Picks the rows of $\boldsymbol{y}$ indicated by the vector $\boldsymbol{\alpha}$\\
$\boldsymbol{J}^{[k]}$ & Sign-change matrix of dataset $k$\\
$\boldsymbol{P}$ & Permutation matrix\\
$\mathcal{I}_X$ & The cost function of method X\\
$G(\cdot)$ & Nonlinearity\\
\hline
\end{tabular}
\label{Table:notations}
\end{table}

\subsection{Independent vector model}
IVA was first introduced by \citet{Taesu2006IVA}. One of the biggest motivations for IVA is that it allows us to study multiple data sets simultaneously. Next, we give a comprehensive review of the IVA model. Throughout this section, we assume that the samples are iid. \revision{Also, throughout the rest of the paper reader can find basic notation from Table \ref{Table:notations}}.
 
\begin{figure}[htb]
    \centering
    \includegraphics[width=1\linewidth]{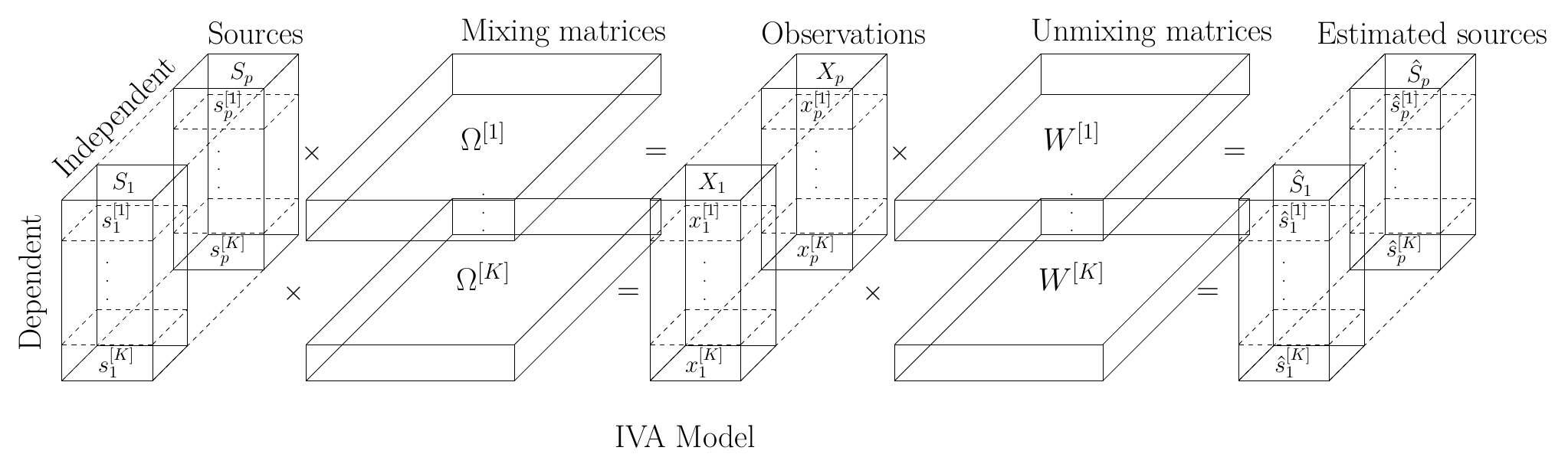}
    \caption{Structure of IVA model with $p$ sources, $p$ observed signals and $K$ data sets. The graph illustrates the mixing process from independent source components to observations and the unmixing process from observations to source estimates.}
    \label{IVAMODELpic}
\end{figure}

Let there be $K$ datasets, each containing $n$ samples, which are formed as a linear mixture of $p$ independent sources. In IVA each data set follows the IC-model \eqref{eq:icmodel}, that is, for $k = 1,\dots,K$,
%Now the $t$th observation of the observed mixture $i$ is denoted as $\boldsymbol{x}_i(t) = (x_i^{[1]}(t),\dots,x_i^{[K]}(t))^{\top}$, where $x_i^{[k]}(t)$ is the observation of the dataset $k$. %Now each observed signal for each dataset $\boldsymbol{x}_1^{[k]},\dots,\boldsymbol{x}_n^{[k]}$ where $k=1,\dots,K$ can be represented as
\begin{align}\label{eq:IVAmodel}
\boldsymbol{x}^{[k]}_i = \boldsymbol{\Omega}^{[k]}\boldsymbol{s}^{[k]}_i + \boldsymbol \mu^{[k]}, \; \; \text{$i = 1,\dots,n$},
\end{align}
where $\boldsymbol{\mu}^{[k]}$ are $p$-variate location (nuisance) parameters, $\boldsymbol{\Omega}^{[k]}$ are full rank $p \times p$ mixing matrices, and $\boldsymbol{s}^{[k]}_1,\dots,\boldsymbol{s}^{[k]}_n$ are independent and identically distributed $p$-variate random vectors. The structure of the IVA model can be seen in the Figure~\ref{IVAMODELpic}.
From now on, we assume that $\boldsymbol{\mu}^{[k]}= \boldsymbol{0}$ for all $k=1,\dots,K$. Before we discuss the assumptions, we give one definition which is crucial for the identifiability of the IVA model.

\begin{definition}[$\boldsymbol{\alpha}$-Gaussianity]
Let $\boldsymbol{\alpha}=(\alpha_1, \dots, \alpha_{K_{\alpha}})^\top \in \mathbb{N}^{K_{\alpha}}$, where $0 \leq K_{\alpha} \leq K$, and write the complement of $\boldsymbol{\alpha}$ as $\boldsymbol{\alpha}^c \in \mathbb{N}^{K-K_{\alpha}}$.
A $K$-vector $\boldsymbol{s}_i=(s_i^{[1]},\dots,s_i^{[K]})^\top$ has an $\boldsymbol{\alpha}$-Gaussian component if $\text{vec}_{\boldsymbol{\alpha}}(\boldsymbol{s}_i)$ and $\text{vec}_{\boldsymbol{\alpha}^{c}}(\boldsymbol{s}_i)$ are independent and $\text{vec}_{\boldsymbol{\alpha}}(\boldsymbol{s}_i) \sim N(\boldsymbol{0}, \boldsymbol{R}_{\boldsymbol{\alpha}})$, where 
$$
\boldsymbol{R}_{\boldsymbol{\alpha}}=\EX(\text{vec}_{\boldsymbol{\alpha}}(\boldsymbol{s}_i)\text{vec}^\top_{\boldsymbol{\alpha}}(\boldsymbol{s}_i))
$$ 
is nonsingular. Above $\text{vec}_{\boldsymbol{\alpha}}(\boldsymbol{s}_i)$ picks the rows of $\boldsymbol{s}_i$ indicated by $\boldsymbol{\alpha}$. 
\end{definition}

Now, the IVA model assumes that

\begin{comment}
IVA model can be represented in matrix form: 
\begin{align*}\    
\boldsymbol{X} = \boldsymbol{\Omega}\boldsymbol{S},
\end{align*}
where $\boldsymbol{\Omega} = \oplus \sum_{k=1}^K \boldsymbol{\Omega}^{[k]}$ and 
$\boldsymbol{S} = \left[ \left(\boldsymbol{S}^{[1]}\right)^{\top},\cdots, \left(\boldsymbol{S}^{[K]}\right)^{\top} \right]^{\top} \in 
\mathbb{R}^{pK \times n}$.

The sources can be also presented as $K \times n$ source component matrices
$\boldsymbol{S}_j = \left[\boldsymbol{s}_j^{[1]}, \dots , \boldsymbol{s}_j^{[K]}\right]^{\top} \in \mathbb{R}^{K\times n}$, $j = 1,..,p$, where 
$\boldsymbol{s}_j^{[k]} = \left[ s_j^{[k]}(1),\dots, s_j^{[k]}(n) \right] \in \mathbb{R}^n$.
\end{comment}

%The mixing matrices $\boldsymbol{\Omega}^{[k]}$ and sources $\boldsymbol{s}^{[k]}(i)$ are unknown quantities to be %estimated only using observed values $\boldsymbol{x}^{[k]}(i)$ and assumptions:
\begin{itemize}
\item[(B1)] The components $s_{i1}^{[k]},\dots,s_{ip}^{[k]}$ are independent for all $k = 1, \dots,K$.
%$\boldsymbol{s}_j^{[k]} \indep \boldsymbol{s}_i^{[k]}$  for all $i \neq j$
\item[(B2)] $\EX(\boldsymbol{s}_i^{[k]})=\boldsymbol 0$ and $\EX(\boldsymbol{s}_i^{[k]}(\boldsymbol{s}_i^{[k]})^\top) = \boldsymbol{I}_p$ for all $k=1,\dots,K$.
\item[(B3)] The components $s_{ij}^{[1]},\dots,s_{ij}^{[K]}$ are dependent on each other for all $i=1,\dots,n$, $j=1,\dots, p$.
%Rows in $\boldsymbol{S}_j, j=1,\dots,p$ are dependent on each other. The source component vectors $\boldsymbol{s}_i \indep \boldsymbol{s}_j$ are $\boldsymbol{\hat{s}}_j=(\hat{s}_j^{[1]},\dots,\hat{s}_j^{[K]})$
\item[(B4)] There are no $K$-vectors $\boldsymbol{s}_l$ and $\boldsymbol{s}_j$ such that $l \neq j$, which have $\boldsymbol{\alpha}$-Gaussian components and 
\begin{align*}
    \boldsymbol{R}_{l, \boldsymbol{\alpha}} = \boldsymbol{D}\boldsymbol{R}_{j, \boldsymbol{\alpha}}\boldsymbol{D},
\end{align*}
where $\boldsymbol{R}_{l, \boldsymbol{\alpha}} = \EX(\text{vec}_{\boldsymbol{\alpha}}(\boldsymbol{s}_l)\text{vec}^\top_{\boldsymbol{\alpha}}(\boldsymbol{s}_l))$ and $\boldsymbol{R}_{j, \boldsymbol{\alpha}} = \EX(\text{vec}_{\boldsymbol{\alpha}}(\boldsymbol{s}_j)\text{vec}^\top_{\boldsymbol{\alpha}}(\boldsymbol{s}_j))$
for any full rank diagonal matrix $\boldsymbol{D} \in \mathbb{R}^{K_{\alpha} \times K_{\alpha} }$.
\end{itemize}

The assumption (B1) is natural because one tries to identify independent components. If one only has the assumptions (B1) and (B4) then the independent components in the model \eqref{eq:IVAmodel} can be recovered only up to scales, signs and permutation, but the arbitrary order of the sources must be the same for all datasets.
Like in ICA, assumption (B2) also fixes the scaling ambiguity of the problem. The assumption (B3) is the new information that can be exploited, and it is the main motivation why \citet{Taesu2006IVA} generalized ICA to IVA in the first place. 
The assumption (B4) is needed for the identifiability of the sources \citep{anderson2013}. One can think of assumption (B4) as an extension of the assumption (A3) in ICA for multiple datasets with dependency across the sources. 

The goal of IVA is to find $K$ unmixing matrices $\boldsymbol{W}^{[1]}, \dots, \boldsymbol{W}^{[K]}$ as well as the corresponding source estimates $\boldsymbol{s}_i^{[k]}$ for each data set such that they all have the same permutation for the sources. The source vector estimates are recovered by
\begin{align*}
    \boldsymbol{s}_i^{[k]} = \boldsymbol{W}^{[k]}\boldsymbol{x}_i^{[k]},  \; i=1,\dots,n, \; k=1,\dots,K.
\end{align*}

In the special case with $K = 1$, IVA reduces to ICA as $\boldsymbol{s}_i$ and $\boldsymbol{s}_j$ reduce to $s_i, s_j \in \mathbb{R}$. 
%Now, assumption (B4) reduces to the assumption (A3). 
In the non-iid case, the identification conditions for IVA 
%and ICA do not change that much from the iid case. The identification conditions 
can be found in \citet[][Theorem 1 and Theorem 2]{anderson2013}.  

\begin{comment}
Note that for independent and identically distributed (iid) samples, i.e. $n=1$, the only change from the above identification condition is that $\boldsymbol{R}_{i,\alpha}=\boldsymbol{D}\boldsymbol{R}_{j,\alpha}\boldsymbol{D}$ instead of $\boldsymbol{R}_{i,\alpha}=(\boldsymbol{I}_n \otimes \boldsymbol{D})\boldsymbol{R}_{j,\alpha}(\boldsymbol{I}_n \otimes \boldsymbol{D})$. From the IVA identification condition one can also get the ICA identification conditions for iid and not iid cases. 

If there is only one data set meaning that $K=1$ and one takes the sample dependency account then $\boldsymbol{S}_i$ and $\boldsymbol{S}_j$ reduce to $\boldsymbol{s}_i, \boldsymbol{s}_j \in \mathbb{R}^n$ respectively, and the diagonal matrix $\boldsymbol{D}$ reduces to real number $\delta$. Now the condition for covariance matrices $\boldsymbol{R}_{i} = \EX(\boldsymbol{s}_i\boldsymbol{s}_i^{\top})$ and $\boldsymbol{R}_{j} = \EX(\boldsymbol{s}_j\boldsymbol{s}_j^{\top})$ reduce to 
\begin{align*}
\boldsymbol{R}_{i} &= (\boldsymbol{I}_n \otimes \boldsymbol{D})\boldsymbol{R}_{j}(\boldsymbol{I}_n \otimes \boldsymbol{D}) \\
& = (\boldsymbol{I}_n \otimes \delta)\boldsymbol{R}_{j}(\boldsymbol{I}_n \otimes \delta)\\
&= (\boldsymbol{I}_n\cdot\delta)\boldsymbol{R}_{j}(\boldsymbol{I}_n \cdot \delta)\\
&= \delta^2\boldsymbol{R}_{j}, \; \; \text{where, $\delta \neq 0$}.
\end{align*} 
\end{comment}

\begin{comment}
    \textcolor{red}{Motivate this example. What do you want to illustrate with it? Maybe no need to write down vectors $\boldsymbol s_i$ and $\boldsymbol s_j$?}

\end{comment}

Next, we give an example of violation of assumption (B4) and show how this leads to violation of assumption (A3) in ICA case. 

\begin{example}
Let $K=5$ and $p = 3$ be the number of data sets and sources, respectively. Now the sources are non-identifiable if there are two source component vectors $\boldsymbol{s}_l$ and $\boldsymbol{s}_j$
\begin{comment}
\begin{equation*}
    \boldsymbol{s}_l = 
        \begin{bmatrix}
            s_l^{[1]}\\
            s_l^{[2]}\\
            s_l^{[3]}\\
            s_l^{[4]}\\
            s_l^{[5]}
        \end{bmatrix}, \; \;        
    \boldsymbol{s}_j =
        \begin{bmatrix}
            s_j^{[1]}\\
            s_j^{[2]}\\
            s_j^{[3]}\\
            s_j^{[4]}\\
            s_j^{[5]}
        \end{bmatrix}, \; \; 
\end{equation*}
\end{comment}
such that $l \neq j$, $l,j = 1,2,3$ and if they have $\boldsymbol{\alpha}$-Gaussian components for which $\boldsymbol{R}_{l,\alpha}=\boldsymbol{D}\boldsymbol{R}_{j,\alpha}\boldsymbol{D}$. This means that from $\boldsymbol{s}_j$ and $\boldsymbol{s}_l$ one can find the same subset of rows, for example $\boldsymbol{\alpha} = [2,3,4]$ for which $\text{vec}_{\boldsymbol{\alpha}}(\boldsymbol{s}_l) \indep \text{vec}_{\boldsymbol{\alpha}^c}(\boldsymbol{s}_l)$, $\text{vec}_{\boldsymbol{\alpha}}(\boldsymbol{s}_i) \sim N(\boldsymbol{0}, \boldsymbol{R}_{l,\boldsymbol\alpha})$ and $\text{vec}_{\boldsymbol{\alpha}}(\boldsymbol{s}_j) \indep \text{vec}_{\boldsymbol{\alpha}^c}(\boldsymbol{s}_j)$, $\text{vec}_{\boldsymbol{\alpha}}(\boldsymbol{s}_j) \sim N(\boldsymbol{0}, \boldsymbol{R}_{j,\boldsymbol\alpha})$. In this case 
\begin{align*}
&\text{vec}_{\boldsymbol{\alpha}}(\boldsymbol{s}_l) = (s_l^{[2]}, s_l^{[3]}, s_l^{[4]})^{\top} \sim N(\boldsymbol{0}, \boldsymbol{R}_{l,\alpha}),\\
&\text{vec}_{\boldsymbol{\alpha}}(\boldsymbol{s}_j) = (s_j^{[2]}, s_j^{[3]}, s_j^{[4]})^{\top} \sim N(\boldsymbol{0}, \boldsymbol{R}_{j,\alpha})
\end{align*}
and $\text{vec}_{\boldsymbol{\alpha}^c}(\boldsymbol{s}_l) = (s_l^{[1]},s_l^{[5]})^\top$, 
$\text{vec}_{\boldsymbol{\alpha}^c}(\boldsymbol{s}_j) = (s_j^{[1]},s_j^{[5]})^\top$.

Notice that if $K=1$ then $\boldsymbol{\alpha} = 1$, $\text{vec}_{\boldsymbol{\alpha}}(\boldsymbol{s}_l) = s_l$ and $\text{vec}_{\boldsymbol{\alpha}}(\boldsymbol{s}_j) = s_j$. Now, assumption (B4) gives us that $s_l \sim N(0,\EX(s_ls_l))$ and $s_j \sim N(0,\EX(s_js_j))$, which means that there are at least two components following normal distributions. This violates the assumption (A3).
\end{example}

\citet{Lahat2016JISA} proved another identification condition for IVA using second-order statistics. Their proof is based on direct factorization of a closed-form expression for the Fisher information matrix. 
%In the paper they proved when the matrix
%\begin{equation*}
%        \begin{bmatrix}
%            \EX(\boldsymbol{s}_j(t)(\boldsymbol{s}_j(t))^{\top}) \odot  \EX(\boldsymbol{s}%_l(t)(\boldsymbol{s}_l(t))^{\top})^{-1}& \boldsymbol{I}\\
%            \boldsymbol{I} & \EX(\boldsymbol{s}_j(t)(\boldsymbol{s}_j(t))^{\top})^{-1} \odot  \EX(\boldsymbol{s}_l(t)(\boldsymbol{s}_l(t))^{\top})
%        \end{bmatrix}
%\end{equation*}\\
%is positive definite, where 
% $\EX(\boldsymbol{s}_l(t)(\boldsymbol{s}_l(t))^{\top}),\EX(\boldsymbol{s}_j(t)(\boldsymbol{s}%_j(t))^{\top})  \in \mathbb{R}^{K \times K}$. 
 %\textcolor{red}{(Are the details needed here, if, check notations for consistency. We don't use $s(t)$ anywhere. )}\\
Further, \citet{Via2011} showed necessary and sufficient identifiability conditions for joint blind source separation using second-order statistics. They proved the identifiability conditions using the idea of equivalently distributed sets of sources. 
%IVA itself is an example of a method to solve the joint blind source separation problem. 

Similarly to ICA, the ambiguities for the IVA model follow from the fact that $\boldsymbol{\Omega}^{[k]}$ and $\boldsymbol{s}_i^{[k]}$ are unknown for all $k = 1,\dots, K$. As IVA is a multidimensional extension of ICA, IVA has the same ambiguities as ICA, but dataset-wise. The mixing matrices $\boldsymbol{\Omega}^{[k]}$, $k=1,\dots,K$ can be estimated up to sign of rows and arbitrary permutation. This means that the sources $\boldsymbol{s}_i^{[k]}$ are considered identified if the mixing matrix $\boldsymbol{\Omega}^{[k]}$ is identified up to $\boldsymbol{\Omega}^{[k]}\boldsymbol{J}^{[k]}\boldsymbol{P}$, where $\boldsymbol{J}^{[k]}$ is sign-change matrix and $\boldsymbol{P}$ is an arbitrary permutation matrix that is common to all datasets. The last statement shows that the IVA model \eqref{eq:IVAmodel} is not unique. 

The first preprocessing step usually in IVA is also data whitening as in ICA. Data whitening is performed for each data set separately \citep{taesu2006}. For some algorithms, whitening is not necessary, but it can ease computations \citep{Anderson2012}.

Lastly, the IVA problem can be solved as several ICA problems by applying ICA to each dataset $k=1,\dots,K$. However, in such approach the information about the dependency assumption is not used which leads to nonoptimal solutions. Also, the permutation of the source estimates might be different for each dataset, as stated in ambiguities of ICA, causing a clustering problem of solving which estimates belong together between the datasets.

\subsection{Cost function and model identification}\label{IVAcost}
To solve the IVA problem, an objective function has to be defined to measure
how good the solution is. The IVA cost function can be derived using mutual information with Kullback-Leibler divergence or entropy like in ICA \citep{Anderson2012}. In the IVA cost function $\mathcal{H}(\boldsymbol{\hat{s}}_j)$ denotes the entropy of the $j$th estimated source component vector \revision{(SCV)} $\boldsymbol{\hat{s}}_j=(\hat{s}_j^{[1]},\dots,\hat{s}_j^{[K]}) =((\boldsymbol{w}_j^{[1]})^{\top}\boldsymbol{x}^{[1]},\dots,(\boldsymbol{w}_j^{[K]})^{\top}\boldsymbol{x}^{[K]})$. From now on index $i$ is dropped, because the source estimates are considered as population-level variables. Using Kullback-Leibler divergence the cost function $\mathcal{I}_{IVA}$ gets the form of 
\begin{align}
    \mathcal{I}_{IVA} &= D_{KL}(p_{\boldsymbol{s}}(\hat{\boldsymbol{s}}_1,\dots,\hat{\boldsymbol{s}}_p)|\prod_{j=1}^p p_{s_j}(\hat{\boldsymbol{s}}_j)) \nonumber \\
                      &=\sum_{j=1}^{p}\mathcal{H}(\boldsymbol{\hat{s}}_j)
                        - \sum_{k=1}^K\log|\det(\boldsymbol{W}^{[k]})| - C \nonumber\\
                      &=\sum_{j=1}^{p}\left( \sum_{k=1}^K\mathcal{H}({\hat{s}}_j^{[k]})-  \mathcal{I}(\boldsymbol{\hat{s}}_j)\right) - \sum_{k=1}^K\log|\det(\boldsymbol{W}^{[k]})| - C \nonumber\\
                      &=\sum_{j=1}^{p}\left( \sum_{k=1}^K\mathcal{H}((\boldsymbol{w}_j^{[k]})^{\top}\boldsymbol{x}^{[k]})-  \mathcal{I}((\boldsymbol{w}_j^{[1]})^{\top}\boldsymbol{x}_j^{[1]},\dots,(\boldsymbol{w}_j^{[K]})^{\top}\boldsymbol{x}_j^{[K]})\right) \nonumber \\
                      &\hspace{0.45cm}- \sum_{k=1}^K\log|\det(\boldsymbol{W}^{[k]})| - C \nonumber,         
\end{align}
%\textcolor{olive}{Should the notations be introduced before the long equation for better readability.}
%\textcolor{red}{Would it be more natural to write this without $\hat s_j$'s, so by replacing it with $Wx_j$? We optimize cost to find W, but $\hat s_j$ already assume it to be found?}
where $C$ is a constant with respect to each unmixing matrix $\boldsymbol{W}^{[k]}$. One can see that if there is no dependency between the sources, then $\mathcal{I}(\hat{\boldsymbol{s}}_j)=0$ for all $j=1, \dots,p$ and the cost function is the same as the sum of $K$ separate ICAs performed for each data set. So, the term $\sum_{j=1}^{p}\mathcal{I}(\boldsymbol{\hat{s}}_j)$ takes the dependency of the sources into account. \revision{From the representation of the cost function one can see that by minimizing $\mathcal{I}_{IVA}$ we simultaneously minimize the entropy of all components and maximize the mutual information within each SCV.} In the case where only one data set is available, that is, $K=1$, then the \revision{SCV} is a scalar quantity, and the IVA cost function reduces to ICA cost function (\ref{ICAcost}). 

\begin{figure}
    \centering
    \includegraphics[width=1\linewidth]{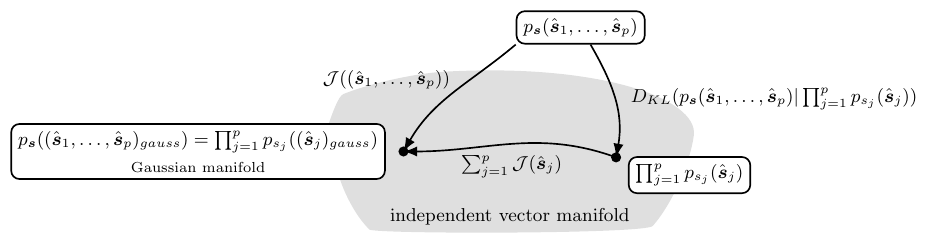}
    \caption{\revision{Visualization of the cost function in the information geometry of zero-mean white probability distributions.}}
    \label{informgeomfigure}
\end{figure}

\revision{As in ICA the IVA cost function can also be linked to maximum likelihood. For detailed explanation how these are linked see \citet[][Section 2.3]{anderson2013thesis}. Another interpretation can be seen from the side of information geometry. For simplicity, let us assume that $\det(\boldsymbol{W}^{[k]}) = 1$ and $\EX(\boldsymbol{s}_j\boldsymbol{s}_j^\top) = \boldsymbol{I}$. As in ICA, negentropy can be used as a cost function in IVA with multidimensional variables. By the assumptions above and Pythagorean theorem in information geometry we get
\begin{align*}
    \mathcal{J}(\hat{\boldsymbol{s}}_j) = -\mathcal{H}(\hat{\boldsymbol{s}}_j),
\end{align*}
where $\mathcal{J}(\hat{\boldsymbol{s}}_j)$ is the negentropy of $\hat{\boldsymbol{s}}_j$. The expression above gives us therefore 
\begin{align*}
    \underset{\boldsymbol{W}^{[1]}, \dots , \boldsymbol{W}^{[K]}}{\operatorname{argmin}} \sum_{j=1}^p \mathcal{H}(\hat{\boldsymbol{s}}_j) = \underset{\boldsymbol{W}^{[1]}, \dots , \boldsymbol{W}^{[K]}}{\operatorname{argmax}} \sum_{j=1}^p \mathcal{J}(\hat{\boldsymbol{s}}_j).
\end{align*}
This relation above can be visualised with information geometry. Information geometry studies statistical manifolds. Statistical manifolds are Riemannian manifolds where each probability density function (pdf) corresponds to a point in the space. To see the connection first define Gaussian manifold as the set of multivariate Gaussian pdfs and independent vector manifold as the set of pdfs where the $\hat{\boldsymbol{s}}_j$ are mutually independent. Now the mutual information and the sum of negentropies can be regarded as distances from $p_{\boldsymbol{s}}(\hat{\boldsymbol{s}}_1,\dots,\hat{\boldsymbol{s}}_p)$ to the Gaussian manifold and independent vector manifold, respectively. Figure \ref{informgeomfigure} provides a schematic view of these relationships. The Gaussian manifold is a single point because zero-mean and white multidimensional Gaussian pdf is unique. The $D_{KL}$ can be seen as projection from $p_{\boldsymbol{s}}(\hat{\boldsymbol{s}}_1,\dots,\hat{\boldsymbol{s}}_p)$ to $\prod_{j=1}^p p_{s_j}(\hat{\boldsymbol{s}}_j)$. Now the goal when updating unmixing matrices is either minimizing the distance 
\begin{align*}
    D_{KL}(p_{\boldsymbol{s}}(\hat{\boldsymbol{s}}_1,\dots,\hat{\boldsymbol{s}}_p)|\prod_{j=1}^p p_{s_j}(\hat{\boldsymbol{s}}_j))
\end{align*}
or maximizing the distance
\begin{align*}
    \sum_{j=1}^p \mathcal{J}(\hat{\boldsymbol{s}}_j).
\end{align*}
}

The derivative of the cost function with respect to each unmixing matrix $\boldsymbol{W}^{[k]}$ gets the form
\begin{align}
    \frac{\partial\mathcal{I}_{IVA}}{\partial\boldsymbol{W}^{[k]}} &= 
    \sum_{j=1}^{p} \frac{\partial\mathcal{H}(\boldsymbol{\hat{s}}_j)}{\partial\boldsymbol{W}^{[k]}}-\frac{\partial\log|\det(\boldsymbol{W}^{[k]})|}{\partial\boldsymbol{W}^{[k]}}\nonumber \\
    &= - \sum_{j=1}^{p} \EX\left( \frac{\partial \log(p(\boldsymbol{\hat{s}}_j))}{\partial\hat{s}_j^{[k]}}\frac{\partial\hat{s}_j^{[k]}}{\partial\boldsymbol{W}^{[k]}} \right) - \left(\left(\boldsymbol{W}^{[k]}\right)^{\top}\right)^{-1}\nonumber \\
    &= \EX\left( \phi^{[k]}(\boldsymbol{\hat{s}}) (\boldsymbol{x}^{[k]})^\top\right) - \left(\left(\boldsymbol{W}^{[k]}\right)^{\top}\right)^{-1} \label{IVAgradient},
\end{align}

where $\phi^{[k]}(\boldsymbol{\hat{s}}) = (\phi^{[k]}(\boldsymbol{\hat{s}}_1),\dots,\phi^{[k]}(\boldsymbol{\hat{s}}_p))^\top$ is formed by selecting $k$th entries from each of the $p$ multivariate score functions, $\phi(\boldsymbol{\hat{s}}_j)=-\partial\log p(\boldsymbol{\hat{s}}_j)/\partial \boldsymbol{\hat{s}}_j$. \revision{In practice, expected values $\EX(\cdot)$ are estimated by sample means.} The last equality comes from the fact that
\begin{align*}
    \left( \frac{\partial\hat{s}_j^{[k]}}{\partial\boldsymbol{W}^{[k]}} \right)_{l,m} = 
    \frac{\partial (\boldsymbol{w}_j^{[k]})^\top \boldsymbol{x}^{[k]}}{\partial w_{l,m}^{[k]}}= x_m^{[k]}\delta_{l,j},
\end{align*}
where $\delta_{l,j}$ denotes the Kronecker delta.
Now, the term $\partial \log(p(\boldsymbol{\hat{s}}_j))/\partial\hat{s}_j^{[k]}$ depends on the assumed distribution of the source components. Notice that the derivative gets a different form if one takes the derivative with respect to the rows of $\boldsymbol{W}^{[k]}$ rather than with respect to $\boldsymbol{W}^{[k]}$, which is done in some algorithms.

\subsection{Source density models}\label{source density models}
As mentioned in Section \ref{IVAcost}, a proper source density model must be chosen to model the true sources. The effectiveness of IVA depends on how well the selected source density models match the true source distributions, as well as on the optimization algorithm used (see Section~\ref{sec:est}). Since the original introduction of IVA \citep{Taesu2006IVA}, numerous source density models with varying properties and applications have been used to solve IVA problems.\\

\textbf{Multivariate Laplace distribution}: 
The multivariate Laplace distribution has been used in many IVA papers, for example, the first IVA paper by \citet{Taesu2006IVA}. There are several ways to define a multivariate extension of the univariate Laplace distribution; the variant considered here is
\[
p(\hat{\boldsymbol{s}}_j) = \frac{\det(\boldsymbol{\Sigma}_j)^{-1/2}}{\Gamma\left(\frac{K+1}{2}\right)2^K\pi^{(K-1)/2}}\exp\left(-\left((\hat{\boldsymbol{s}}_j-\boldsymbol{\mu}_j)^{\top}\boldsymbol{\Sigma}_j^{-1}(\hat{\boldsymbol{s}}_j-\boldsymbol{\mu}_j)\right)^{1/2}\right),
\]
where $\hat{\boldsymbol{s}}_j$ is a $K$-variate random vector, %$\in \mathbb{R}^{K}$, 
$\boldsymbol{\mu}_j$ is the $K$-variate location parameter and $\boldsymbol{\Sigma}_j$ is the positive-definite $K \times K$ scatter matrix \citep{Arslan2010}.
The score function for the multivariate Laplace distribution is
\begin{align*}
    \phi(\hat{\boldsymbol{s}}_j) &= - \frac{\partial\log(p(\hat{\boldsymbol{s}}_j))}{\partial \hat{\boldsymbol{s}}_j}\\
    &= \frac{\boldsymbol{\Sigma}_j^{-1}(\hat{\boldsymbol{s}}_j-\boldsymbol{\mu}_j)}{((\hat{\boldsymbol{s}}_j-\boldsymbol{\mu}_j)^\top\boldsymbol{\Sigma}_j^{-1}(\hat{\boldsymbol{s}}_j-\boldsymbol{\mu}_j))^{1/2}}.
\end{align*}\\

\textbf{Gaussian distribution}: IVA with Gaussian source distribution was introduced in \citet{Anderson2010trick}. 
%in which they also presented new IVA algorithms
%Assumes that the source estimates \textcolor{olive}{why should the estimates be Gaussian - this is the assumption of the sources, or?} are multivariate Gaussian distributed. The distribution for the estimated sources $\hat{\boldsymbol{s}}_j$ is
The pdf of the multivariate Gaussian distribution is 
\[
p(\hat{\boldsymbol{s}}_j) = (2\pi)^{-K/2}\det(\boldsymbol{\Sigma}_j)^{-1/2}
\exp\left(-\frac{1}{2}(\hat{\boldsymbol{s}}_j - \boldsymbol{\mu}_j)^{\top}\boldsymbol{\Sigma}_j^{-1}(\hat{\boldsymbol{s}}_j - \boldsymbol{\mu}_j)\right),
\] 
where $\boldsymbol{\mu}_j$ is the $K$-variate location parameter and $\boldsymbol{\Sigma}_j$ is the positive-definite $K \times K$ scatter matrix. 
The multivariate score function $\phi(\hat{\boldsymbol{s}}_j)$ has the form of
\begin{equation*}
    \phi(\hat{\boldsymbol{s}}_j) = \boldsymbol{\Sigma}_j^{-1}(\hat{\boldsymbol{s}}_j-\boldsymbol{\mu}_j).
\end{equation*}\\
%\textcolor{olive}{Should here (or somewhere else) be discussed the nice properties of Gaussian distribution which makes the optimization with decoupling trick super fast compared to other source densities?} \\
\textbf{Multivariate Student’s t-distribution}: The multivariate Student's t-distribution was used in the IVA context first by \citet{Liang2013}. The pdf takes the form of
\[
p({\hat{\boldsymbol{s}}_j}) = \frac{\Gamma\left(\frac{\nu+K}{2}\right)\det(\boldsymbol{\Sigma}_j)^{-1/2}}{\Gamma\left(\frac{\nu}{2}\right)\nu^{K/2}\pi^{K/2}} \left( 1 + \frac{(\hat{\boldsymbol{s}}_j - \boldsymbol{ \mu}_j)^{\top}\boldsymbol{\Sigma}_j^{-1}
(\hat{\boldsymbol{s}}_j - \boldsymbol{\mu}_j)}{\nu}\right)^{-(\nu+K)/2}.
\]
where $\boldsymbol{\mu}_j$ is the $K$-variate location parameter, $\boldsymbol{\Sigma}_j$ is the positive-definite $K \times K$ scatter matrix, $\nu$ represents the degrees of freedom , and $\Gamma(\cdot)$ denotes the Gamma function. When one decreases the parameter $\nu$, the distribution becomes more heavy-tailed. One can get the multivariate Cauchy distribution as a special case with $\nu = 1$. The multivariate score function for Student's t-distribution is
%\textcolor{olive}{Should you make clearer that this is for t and not Cauchy. Also at the end of the equation should be ``.''}
\begin{align*}
     \phi(\hat{\boldsymbol{s}}_j) 
     &= \frac{(\nu + k)\boldsymbol{\Sigma}_j^{-1}(\hat{\boldsymbol{s}}_j-\boldsymbol{\mu}_j)}{\nu + (\hat{\boldsymbol{s}}_j - \boldsymbol{ \mu}_j)^{\top}\boldsymbol{\Sigma}_j^{-1}
(\hat{\boldsymbol{s}}_j - \boldsymbol{\mu}_j)}.
\end{align*}\\

\textbf{Kotz distribution}: The Kotz distribution family was used for IVA in \citet{Anderson2013kotz}. The pdf of a $K$-dimensional Kotz distribution is: 
\begin{align*}
    p({\hat{\boldsymbol{s}}_j})=&{{\Gamma}\left(\frac{K}{2}\right)\beta\lambda^{\nu}\det{(\boldsymbol{\Sigma}}_j)^{-1/2}({(\hat{\boldsymbol{s}}_j}-\boldsymbol{\mu})^{\top}{\boldsymbol{\Sigma}_j}^{-1} ({\hat{\boldsymbol{s}}_j}-\boldsymbol{\mu}))^{\eta-1}\over {\Gamma}(\nu)\pi^{K/2}}\\
    &\times\exp(-\lambda(({\hat{\boldsymbol{s}}_j}-\boldsymbol{\mu})^{\top}{{\boldsymbol{\Sigma}_j}^{-1}{(\hat{\boldsymbol{s}}_j}-\boldsymbol{\mu}))^{\beta}}),
\end{align*}
where $\boldsymbol{\theta} = [\beta, \eta, \lambda]^{\top}$ are the scalar parameters for Kotz distribution, namely, $\lambda>0$ (kurtosis parameter), $\beta>0$ (shape parameter), $\eta>(2 - K)/2$, $\boldsymbol{\mu}_j$ is the $K$-variate location parameter, $\boldsymbol{\Sigma}_j$ is positive-definite $K \times K$ scatter matrix and $\nu = (2\eta + K - 2)/(2\beta) > 0$. The multivariate score function for the Kotz distribution is given by
\begin{equation*}
    \phi(\hat{\boldsymbol{s}}_j) = 2\left(1 - \eta + \lambda \beta \left((\hat{\boldsymbol{s}}_j - \boldsymbol{\mu})^{\top} \boldsymbol{\Sigma}_j^{-1} (\hat{\boldsymbol{s}}_j - \boldsymbol{\mu})\right)^{\beta} \right) 
    \frac{\boldsymbol{\Sigma}_j^{-1} (\hat{\boldsymbol{s}}_j - \boldsymbol{\mu})}
         {(\hat{\boldsymbol{s}}_j - \boldsymbol{\mu})^{\top} \boldsymbol{\Sigma}_j^{-1} (\hat{\boldsymbol{s}}_j - \boldsymbol{\mu})}.
\end{equation*}
By selecting the parameter vector $\boldsymbol{\theta} = [\beta, \eta, \lambda]^{\top}$ appropriately, one can recover the score functions of various distributions. For example, setting $\boldsymbol{\theta} = [1, 1, 1/2]^{\top}$ yields the score function of the multivariate Gaussian distribution, whereas $\boldsymbol{\theta} = [1/2, 1, 1]^{\top}$ results in the score function for the multivariate Laplace distribution.

The Kotz distribution also encompasses another distribution family known as the multivariate power exponential (MPE) family~\citep{Gomez1998}. The pdf for the $K$-dimensional MPE family is
\begin{align*}
    p(\hat{\boldsymbol{s}}_j) = \frac{\Gamma\left(\frac{K}{2}\right)K\det(\boldsymbol{\Sigma}_j)^{-1/2}}{  \Gamma\left(1 + \frac{K}{2\beta}\right)\pi^{1/2}2^{1 + \frac{K}{2\beta}}}\exp(-\frac{1}{2}((\hat{\boldsymbol{s}}_j-\boldsymbol{\mu}_j)^{\top}\boldsymbol{\Sigma}_j^{-1}(\hat{\boldsymbol{s}}_j - \boldsymbol{\mu}_j))^{\beta}),
\end{align*}
where $\beta \in (0, \infty)$, $\boldsymbol{\mu}_j$ is the $K$-variate location parameter and $\boldsymbol{\Sigma}_j$ is positive-definite $K \times K$ scatter matrix. If one selects $\lambda = \frac{1}{2}$ and $\eta = 1$ then the Kotz distribution reduces to the  MPE distribution after some simple algebra.\\

\textbf{Multivariate generalized Gaussian distribution \revision{(MGGD)}}: In IVA \revision{MGGD} have been used in \cite{LIANG2014175,Boukouvalas2015}. The density for the MGGD family has the form
\[
p(\hat{\boldsymbol{s}}_j) = \frac{\Gamma\left(\frac{K}{2}\right)\beta\det(\boldsymbol{\Sigma}_j)^{-1/2}}{\Gamma\left(\frac{K}{2\beta}\right)\pi^{K/2}\alpha^{K/2}} \exp\left(-\left(\frac{(\hat{\boldsymbol{s}}_j - \boldsymbol{\mu}_j)^{\top}\boldsymbol{\Sigma}_j^{-1}
(\hat{\boldsymbol{s}}_j - \boldsymbol{\mu}_j)}{\alpha}\right)^{\beta} \right),
\]
where $\boldsymbol{\mu}_j$ is the $K$-variate location parameter, $\boldsymbol{\Sigma}_j$ is positive-definite $K \times K$ scatter matrix, and $\alpha, \beta \in \mathbb{R}^+$ are the scale and shape parameters respectively. Note that by choosing $\alpha = 1$ and $\beta = \frac{1}{2}$ one gets the multivariate Laplace distribution above, and $\alpha=2$ and $\beta=1$ gives the Gaussian distribution. The multivariate score function is 
\begin{align*}
    \phi(\hat{\boldsymbol{s}}_j) 
    &= \frac{2\beta((\hat{\boldsymbol{s}}_j - \boldsymbol{\mu}_j)^{\top}\boldsymbol{\Sigma}_j^{-1}
(\hat{\boldsymbol{s}}_j - \boldsymbol{\mu}_j))^{\beta-1}}{\alpha^{\beta}}\boldsymbol{\Sigma}_j^{-1}(\hat{\boldsymbol{s}}_j-\boldsymbol{\mu}). 
\end{align*}
\\

\textbf{Mixed distribution}: In some situations, the true sources are not from one distribution, but rather from a mixed distribution. \citet{Rafique2016} proposed a mixed source distribution of the multivariate Student's t and multivariate Laplace distribution. The density of the mixed distribution is
\[
p(\hat{\boldsymbol{s}}_j) = \epsilon f_{St} + (1-\epsilon)f_L,
\]
where $f_{St}$ and $f_L$ are the multivariate Student's t-distribution and multivariate Laplace distribution, respectively, and $\epsilon \in [0,1]$ is a weighting parameter, which determines the weight of each distribution. The multivariate score function is the sum of the score functions of $f_{St}$ and $f_L$ multiplied by the weighting parameters.\\

\textbf{Super-Gaussian distribution}: In IVA super-Gaussian distribution gained popularity after the paper by \citet{Ono2011}. If random variable $\hat{\boldsymbol{s}}_j$ has probability density of the form
\begin{align*}
    p(\hat{\boldsymbol{s}}_j) = \exp(-G_R(\hat{\boldsymbol{s}}_j)),
\end{align*}
where $G_R(\hat{\boldsymbol{s}}_j)$ is an even function, which is differentiable, except possibly at the origin, $G_R(\hat{\boldsymbol{s}}_j)$ strictly increasing in $\mathbb{R}_+$, and $G'_R(\hat{\boldsymbol{s}}_j)/\hat{\boldsymbol{s}}_j$ is strictly decreasing in $\mathbb{R}_+$, then it is called super-Gaussian \citep[p.60-61]{benveniste2012adaptive}. The multivariate score function is then
\begin{align*}
     \phi(\hat{\boldsymbol{s}}_j)
    &=\frac{\partial G_R(\hat{\boldsymbol{s}}_j)}{\partial \hat{\boldsymbol{s}}_j}.
\end{align*}
\\

%\textbf{Circular Gaussian distribution}: Give definition and cite FastDIVA paper

\begin{comment}
    \textbf{Deep neural network}: \citet{Li2020IndependentVA} proposed deep neural network source priors with natural gradient descent as an optimizer for the speech separation task. \citet{Kang2021} proposed a real-time blind source separation algorithm, which used a deep neural network source prior with auxiliary-function based IVA. (Move this to discussion)
\end{comment}

In the next section, we review some common estimation methods used to solve the IVA problem. 

\subsection{Estimation methods}\label{sec:est}

Popular choices for the optimization method that solve the IVA problem are gradient descent, natural gradient descent, Newton's method and block Newton's method, for example. In the following section, we go through some of the most common algorithms. A detailed overview of optimization methods can be found in \citet{Guo2023review}. \revision{Notice that all of the methods here assume the IVA assumptions (B1)–(B4), but in some methods (B4) is replaced by a stronger assumption by requiring that latent components come from specific distribution family.}\\

\textbf{Matrix (natural) gradient descent}: The simplest method to solve the IVA problem is the (natural) gradient descent algorithm. Each of the $K$ matrices is then updated sequentially using the steepest descent method
\begin{align} \label{Gradient update}
    \boldsymbol{W}^{[k]} \leftarrow \boldsymbol{W}^{[k]} - \rho\frac{\partial\mathcal{I}_{IVA}}{\partial \boldsymbol{W}^{[k]}},
\end{align}
where $\rho$ is the step size, which can be fixed to be any positive value between zero and one. It is suggested to use natural gradient update for faster convergence. The natural gradient is the gradient of (\ref{IVAgradient}) postmultiplied by $(\boldsymbol{W}^{[k]})^\top \boldsymbol{W}^{[k]}$ and the update is now given by
\begin{align*}
    \boldsymbol{W}^{[k]} \leftarrow \boldsymbol{W}^{[k]}- \rho\frac{\partial\mathcal{I}_{IVA}}{\partial \boldsymbol{W}^{[k]}}(\boldsymbol{W}^{[k]})^\top \boldsymbol{W}^{[k]}.
\end{align*}
\revision{Out of all methods we give in this section, the natural gradient converges slowest of them all. This previous statement has been empirically verified in several simulation studies, see for example \citet{lee2007,Ono2011,Anderson2012}. Even if the computational complexity of natural gradient is low, it is advised to use other methods that converge faster.}\\

\textbf{Newton update IVA}: 
When considering the update rule (\ref{Gradient update}) or its natural gradient version, the main issue lies in using a single step size to update the entire unmixing matrix. This approach assigns equal weight to all directions, which can be undesirable. \citet{Anderson2010trick} proposed to sequentially update each row of the unmixing matrix via a decoupling method. This decoupling method allows one to change the step size for each direction, which gives faster \revision{convergence} compared to matrix gradient and matrix natural gradient.
%One of the simplest algorithms for IVA is to use non-orthogonal Newton's update. Algorithm uses Newton update with a decoupling trick from \citet{Anderson2010trick} to update the unmixing matrices. 

In the decoupling trick, one defines a unit length vector $\boldsymbol{h}_j^{[k]}$ such that $\tilde{\boldsymbol{W}}_j^{[k]}\boldsymbol{h}_j^{[k]}=\boldsymbol{0}$, where $\tilde{\boldsymbol{W}}_j^{[k]}$ is a $(p-1)\times p$ matrix formed by removing the $j$th row from $\boldsymbol{W}^{[k]}$. \citet{Li2007} showed that
\begin{align*}
    |\det(\boldsymbol{W}^{[k]})| = |(\boldsymbol{h}_j^{[k]})^\top \boldsymbol{w}_j^{[k]}|
    \left(|\det(\tilde{\boldsymbol{W}}_j^{[k]}(\tilde{\boldsymbol{W}}_j^{[k]})^\top)|\right)^{1/2}.
\end{align*}
\revision{Let us denote here the unmixing vector that estimates the $j$th SCV as $\boldsymbol{w}_j^\top = ( (\boldsymbol{w}_j^{[1]})^\top, \dots, (\boldsymbol{w}_j^{[K]})^\top)$.} \revision{Then the} update is done using the \revision{$\boldsymbol{w}_j$}
%source component vectors of the unmixing matrices 
rather than the unmixing matrices $\boldsymbol{W}^{[k]}$. The gradient of the IVA cost function with respect to the 
%$j$th source component vector 
$\boldsymbol{w}_j$
%$\boldsymbol{w}_j^\top = ((\boldsymbol{w}_j^{[1]})^\top, \dots,(\boldsymbol{w}_j^{[K]})^\top)$ 
is
\begin{align*}
    \frac{\partial \mathcal{I}_{IVA}}{\partial \boldsymbol{w}_j} = \left( \left( \frac{\partial \mathcal{I}_{IVA}}{\partial \boldsymbol{w}_j^{[1]}} \right)^{\top}
    , \dots, 
    \left( \frac{\partial \mathcal{I}_{IVA}}{\partial \boldsymbol{w}_j^{[K]}}  \right)^\top \right)^{\top},
\end{align*}
and the Hessian matrix $\mathbf{H} = \partial^2 \mathcal{I}_{IVA}/\partial\boldsymbol{w}_j \partial\boldsymbol{w}_j^\top$ is a block matrix with $K$ rows and $K$ columns consisting of $p \times p$ matrices. The block diagonal entries for the Hessian matrix are
\begin{align*}
    \mathbf{H}_{k,k} = \EX\left( \frac{\partial \phi^{[k]}(\boldsymbol{\hat{s}}_j)}{\partial \boldsymbol{w}_j^{[k]}}(\boldsymbol{x}^{[k]})^\top \right) + \frac{\boldsymbol{h}_j^{[k]}(\boldsymbol{h}_j^{[k]})^\top}{((\boldsymbol{h}_j^{[k]})^\top \boldsymbol{w}_j^{[k]})^2}
\end{align*}
and the off-block diagonal entries are 
\begin{align*}
    \mathbf{H}_{k_1,k_2} = \EX \left( \frac{\partial \phi^{[k_2]}(\boldsymbol{\hat{s}}_j)}{\partial \boldsymbol{w}_j^{[k_1]}}(\boldsymbol{x}^{[k_2]})^\top \right),
\end{align*}
where $k_1 \neq k_2$. The Hessian matrix $\mathbf{H}$ can be strictly positive definite or not, but this depends on the assumed distribution for the \revision{SCV}. \citet{Anderson2012} showed that the Hessian matrix is always strictly positive definite for a multivariate Gaussian source prior.

Now, the Newton's method is used to update the 
%source component vectors 
$\boldsymbol{w}_j$ with update rule
\begin{align*}
    \boldsymbol{w}_j \leftarrow \boldsymbol{w}_j - \rho \mathbf{H}^{-1}\frac{\partial \mathcal{I}_{IVA}}{\partial \boldsymbol{w}_j},
\end{align*}
where $\rho$ is the step size, which can be fixed to be any positive value between zero and one. This Newton algorithm described above can be used if the Hessian matrix $\mathbf{H}$ is positive definite. If the Hessian is ill-conditioned, e.g. singular, then variations of Newton’s method can be utilized \citep{anderson2013thesis}.\\

\textbf{Fast fixed-point independent vector analysis (FastIVA)}: FastIVA was first introduced by \citet{lee2007} and they derived the algorithm for the complex source case. The sources were assumed to be circular and unmixing matrices were constrained to be orthogonal. The cost function for FastIVA is
\begin{align*}
    \mathcal{I}_{\text{FastIVA}} =  \sum_{j=1}^p \EX \left( G\left( \sum_{k=1}^K 
    |\hat{s}_j^{[k]}|^2 \right) \right),
\end{align*}
where $G(\cdot)$ is a nonlinearity with the relation of 
\begin{align*}
    G\left( \sum_{k=1}^K |\hat{s}_j^{[k]}|^2 \right) = 
    -\log(\hat{p}_{s_j}(\boldsymbol{\hat{s}}_j)),
\end{align*}
where $\hat{s}_j^{[k]} = (\boldsymbol{w}_j^{[k]})^\top \boldsymbol{x}^{[k]}$ and $\hat{p}_{s_j}(\boldsymbol{\hat{s}}_j)$ denotes the estimated source pdf. 
There are several choices for the function $G$, for example, $G_1(u) = \log(u)$, $G_2(u) = \sqrt{u}$, $G_3(u) = \sqrt{2/K}\sqrt{u} + (K - 1/2)\log(u)$ and $G_4 = \sqrt[3]{u}$. The form for $G_1$ comes from entropy estimation, which was derived for spherically symmetric sources by \citet{Lee2007entropicapprox}. The forms of the functions $G_2$ and $G_3$ come from the pdf choices. The pdf choices used by \citet{lee2007} can be found in \ref{FastIVAapp}. The $G_4$ was derived by \citet{LIANG2014175} using MGGD distribution.
Note that the cost function is the same as above because of the restriction of orthogonal unmixing matrices and because one can also drop the constant term. 

The update rule in FastIVA is
\begin{align*}
    \boldsymbol{w}_j^{[k]} \leftarrow &\EX\left( G'\left( \sum_{k=1}^K 
    |\hat{s}_j^{[k]}|^2 \right) + |\hat{s}_j^{[k]}|^2G''\left( \sum_{k=1}^K 
    |\hat{s}_j^{[k]}|^2 \right) \right)\boldsymbol{w}_j^{[k]} \\
    &- \EX \left( (\hat{s}_j^{[k]})G'\left( \sum_{k=1}^K 
    |\hat{s}_j^{[k]}|^2 \right)\boldsymbol{x}^{[k]} \right),
\end{align*}
where $G'(\cdot)$ and $G''(\cdot)$ denote the derivative and the second derivative of the nonlinear function $G(\cdot)$ respectively.
After updating the matrix $\boldsymbol{W}^{[k]}$, one needs to normalize and decorrelate the rows. In \citet{lee2007} they used symmetrical decorrelation scheme:
\begin{align*}
    \boldsymbol{W}^{[k]} \leftarrow \left( \boldsymbol{W}^{k} (\boldsymbol{W}^{[k]})^\top \right)^{-1/2}\boldsymbol{W}^{[k]}.
\end{align*}
\revision{\citet{Guo2023review} tested different IVA algorithms through the separation of convolutional mixed speech signals. The simulation studies were done in three different settings and FastIVA outperformed the other algorithms in one of these settings. In the other settings FastIVA was still performing well. In the paper it was said that the time complexity of FastIVA is moderate. As a side note \citet{Wang2025} improved FastIVA. Simulation experiments showed that the improved FastIVA iteration count and run time was much smaller when compared to other algorithms in the paper.}

\citet{Qian2014} extended FastIVA to noncircular sources and studied local stability analysis of the FastIVA algorithm. FastIVA has also been generalized by Fast Dynamic Independent Vector Analysis algorithm (FastDIVA) \citep{Koldovsky2021}.\\

\textbf{IVA-G} \citep{Anderson2012}: Assumes multivariate Gaussian distributed sources. The cost function is then 
\begin{align}
    \mathcal{I}_{IVA-G} = \frac{pK\log(2\pi e)}{2} + \frac{1}{2}\log \left( \prod_{j=1}^p \prod_{k=1}^K \lambda_{j,k} \right)
    - \sum_{k=1}^K \log(|\det(\boldsymbol{W}^{[k]})|) - C \label{IVA-Gcost},
\end{align}
where $\lambda_{j,k}$ is the $k$th eigenvalue of the covariance matrix $\boldsymbol{\Sigma}_j= \EX(\boldsymbol{s}_j\boldsymbol{s}_j^\top)$ associated with $j$th \revision{SCV}. The cost function indicates that the product of eigenvalues over all of the source component vectors should be minimized. 

When using multivariate Gaussian distribution assumption for the sources we cannot assume $\boldsymbol{\Sigma}_j$ to be identity matrix because otherwise the assumption that the sources are dependent is not satisfied. This means that $\boldsymbol{\Sigma}_j$ has to be estimated during the optimization process. 

\citet{Anderson2012} proposed four different IVA-G algorithms and compared them with some existing algorithms using simulations. The proposed algorithms were matrix gradient update IVA-G-M, vector gradient update IVA-G-V, Newton update IVA-G-N and block Newton update IVA-G-B. The IVA-G algorithms make use of second-order statistics and \citet{Anderson2012} gave the local stability conditions for IVA-G. As a side note, IVA-G-N has been shown to be computationally efficient algorithm through simulations \citep{Anderson2010trick,anderson2013}. IVA-G-N is particularly fast to compute as compared to the other algorithms. The reason for this is that decoupling trick is very efficient with Gaussian distribution \citep{anderson2013thesis}.

\revision{\citet{Anderson2012} also proposed approach in which one uses two algorithms in succession. They called the approach IVA-GL (Gaussian-Laplace). First one uses, for example, IVA-G-N and then initializes the second algorithm with the IVA-G-N solution. For the second algorithm, one can use, for example, matrix gradient or Newton update, but one needs to assume that the SCV come from a Laplace distribution and that the covariance matrix for every SCV is a scaled identity matrix. More about IVA-GL one can find from \citep{Anderson2012,Engberg2016}.}
\\

\textbf{Auxiliary-function-based IVA (AuxIVA)}: AuxIVA was first introduced by \cite{Ono2011} and is a direct generalization of auxiliary-function-based ICA \citep{Ono2010}. %The AuxIVA algorithm consists of two alternative updates which are 1) weighted covariance matrix update and 2) unmixing matrix update. 
AuxIVA does not include any tuning parameters and the monotonic decrease of the cost function at each update is guaranteed. 

Auxiliary function technique can be considered as an extension of expectation-maximization algorithm. In this technique, one designs a function $Q(\theta, \tilde{\theta})$ such that is satisfies
\[
C(\theta) = \min_{\tilde{\theta}}Q(\theta, \tilde{\theta}),
\]
where $C(\theta)$ is the cost function. The $Q(\theta, \tilde{\theta})$ and $ \tilde{\theta}$ are called the auxiliary function for $C$ and auxiliary variable, respectively. Instead of minimizing the cost function one minimizes the auxiliary function $Q(\theta, \tilde{\theta})$ in terms of $\theta, \tilde{\theta}$ one-by-one as follows 
\begin{align*}
    \tilde{\theta}^{(i+1)} &= \arg \min_{\tilde{\theta}} Q(\theta^{(i)}, \tilde{\theta}) \\
    \theta^{(i+1)} &= \arg \min_{\theta} Q(\theta, \tilde{\theta}^{(i+1)}),
\end{align*}
where $i$ denotes the iteration index.

In IVA case, \citet{Ono2011} derived the auxiliary function and auxiliary variables. They get the forms of
\begin{align*}
    Q(\mathcal{W}, \mathcal{V})&= \sum_{k=1}^K Q_k(\boldsymbol{W}^{[k]}, \boldsymbol{V}^{[k]}),\\
    Q_k(\boldsymbol{W}^{[k]}, \boldsymbol{V}^{[k]})&= \frac{1}{2}\sum_{j=1}^p (\boldsymbol{w}_j^{[k]})^{\top}\boldsymbol{V}_j^{[k]}\boldsymbol{w}_j^{[k]} - \log|\det (\boldsymbol{W}^{[k]})| + R,
\end{align*}
where
\begin{align*}
    \boldsymbol{V}_j^{[k]} = \EX\left(\frac{G_R'(r_j)}{r_j}\boldsymbol{x}^{[k]}(\boldsymbol{x}^{[k]})^{\top}\right),
\end{align*}
and \revision{$\mathcal{W}$ denotes the set $\{\boldsymbol{W}^{[1]},\dots, \boldsymbol{W}^{[K]}\}$,} $r_j$ is a positive random variable, $\boldsymbol{V}^{[k]}$ represents a set of $\boldsymbol{V}_j^{[k]}$ for any $j$, $\mathcal{V}$ \revision{represents} set of $\boldsymbol{V}_j^{[k]}$ for any $j,k$, $R$ is a constant independent of $\boldsymbol{W}^{[k]}$, and $G'_R(r_j)$ is the derivative of spherical function $G_R(r_j)$ introduced in subsection \ref{source density models}.
\begin{comment}
Now the derivative of the auxiliary function with respect to the $\boldsymbol{w}_j^{[k]}$ is
\begin{align*}
    \frac{\partial Q(\mathcal{W}, \mathcal{V})}{\partial \boldsymbol{w}_j^{[k]}} = 
     V_j^{[k]}\boldsymbol{w}_j^{[k]} - \frac{\partial \log|\det \boldsymbol{W}^{[k]}|}{\partial \boldsymbol{w}_j^{[k]}}.
\end{align*}
\end{comment}

The cost function in AuxIVA is
\begin{align*}
    \mathcal{I}_{Aux} = \sum_{j=1}^p\EX(G(\boldsymbol{\hat{\boldsymbol{s}}}_j))-\sum_{k=1}^K|\det(\boldsymbol{W}^{[k]})|,
\end{align*}

where $\hat{\boldsymbol{s}}_j = (\hat{s}_j^{[1]}, \dots, \hat{s}_j^{[K]})^\top$ and nonlinearity function $G(\cdot)$ has a relation of
\begin{align*}
    G(\boldsymbol{\hat{\boldsymbol{s}}}_j)= -\log p(\boldsymbol{\hat{\boldsymbol{s}}}_j).
\end{align*}

\citet{Ono2011} gave two assumptions for the nonlinearity $G$. The first one was spherical symmetry for $G(\boldsymbol{\hat{\boldsymbol{s}}}_j)$ meaning that $G(\boldsymbol{\hat{\boldsymbol{s}}}_j)$ can be represented as %\textcolor{red}{Can you just write $G(||\hat{\boldsymbol s}_j||_2)$, where $||\cdot||_2$ is the Eucledean norm.}
\begin{align*}
    G(\boldsymbol{\hat{\boldsymbol{s}}}_j) &= G_R(||\boldsymbol{\hat{\boldsymbol{s}}}_j||_{2}),
\end{align*}
 where $||\cdot||_2$ is the Euclidean norm.
%where $G_R(r_j)$ is a function of a real-valued scalar variable, $r_j$ and $||\cdot||_{l^2}$ denotes the $l^2$-norm of a vector.
The second assumption was that the sources are from the super Gaussian distribution. After the assumptions, it can be shown \citep[Theorem 2]{Ono2011} that for any $\mathcal{W}$ and $\mathcal{V}$
\begin{align*}
    \mathcal{I}_{Aux} \leq Q(\mathcal{W}, \mathcal{V})
\end{align*}
holds. The update rules for the auxiliary variables are presented in \citet{Ono2011,Ono2012}.
\begin{comment}
The equality sign holds if and only if \textcolor{red}{Why absolute values in the norm below?}\textcolor{blue}{Because it is $l_2$-norm. Forgot to specify it.}\textcolor{red}{Is s complex here then?}
\begin{align*}
    r_j = ||\boldsymbol{\hat{\boldsymbol{s}}}_j||_{2} = \left( \sum_{k=1}^K (\hat{s}_j^{[k]})^2\right)^{1/2} .
\end{align*}
\end{comment}
\revision{Several improved versions of AuxIVA employ alternative strategies for updating the unmixing matrix. For interested readers we refer to \citet{Scheibler2021} and \citet{Guo2023review} for further details.}

\revision{The convergence speed of AuxIVA and its variants have been studied using simulation studies in \citet{Ono2011,Scheibler2021,Guo2023review}. The overall conclusion is that the convergence speed for AuxIVA and its different variants is really fast. However, even if the different variants converge fast their complexities can be quite different. \citet{Guo2023review} compared eight different AuxIVA variants and their time complexities which were there categorized as low, moderate or high.}

AuxIVA has been used in many other developed IVA algorithms. Examples include sparseAuxIVA \citep{Janský2016}, overdetermined IVA \revision{algorithms} \citep[OverIVA,][]{Scheibler2019,Scheibler2021}, geometrically constrained IVA \revision{algorithms} \cite[\revision{GCAV-IVA},][]{Li2020geom}, and FasterIVA \citep{Andreas2021} to name a few. 

In addition to the methods discussed in the paper, there are several other IVA methods, which are not reviewed here. Table \ref{Table-assumptions} summarizes various methods along with the underlying assumptions. In addition, Table \ref{Table-implementations} gives different algorithms and software implementations available in R, Matlab and Python for the listed methods. 
%\textcolor{red}{But real-time IVA and blinky IVA are not mentioned in the table, right?}
%real-time IVA by \citet{Kang2021} and blinky IVA by \citet{Scheibler2019blinky}

\subsection{Performance metrics}
The most used performance metric to evaluate the performance of methods in simulation studies is joint intersymbol interference (jISI) defined by \cite{anderson2013}. The jISI is an extension of intersymbol interference \citep[ISI,][]{Amari1995}. The jISI is defined as
\begin{align*}
    ISI_{joint} = \frac{1}{2p(p-1)}\left( \sum_{s=1}^p 
    \left( \sum_{j=1}^p \frac{\overline{g}_{s,j}}{\max_l \overline{g}_{s,l}} - 1\right) + \sum_{s=1}^p \left( \sum_{j=1}^p \frac{\overline{g}_{s,j}}{\max_l \overline{g}_{l,m}} - 1\right) \right),
\end{align*}
where $\overline{g}_{s,j} = \sum_{k=1}^K|g_{s,j}^{[k]}|$, and $g_{s,j}^{[k]}$ is the entry of the $s$th row and $j$th column from matrix $\boldsymbol{G}^{[k]}=\boldsymbol{W}^{[k]}\boldsymbol{\Omega}^{[k]}$. The joint ISI metric penalizes
source component vector estimates that are not consistently aligned across datasets and it is normalized so that $0 \leq ISI_{joint} \leq 1$, with 0 implying ideal separation performance.

Other performance metrics which can be used in IVA are interference-to-signal power ratio and separation success ratio. Information about these metrics and how they can be used in IVA can be found in \citet{anderson2013thesis}.

\begin{comment}
    Other common choices for the performance metrics are source-to-distortion ratio (SDR), source-to-interferences ratio (SIR), source-to-noise ratio (SNR), sources-to-artifacts ratio (SAR) and signal-to-interference and noise ratio (SINR) for example. Fore more details of these performance metrics, see \citet{Vincent2006}. \textcolor{olive}{is that really a good reference? Isn't that only for standard BSS where a single matrix is to be estimated? Otherwise I have also a review paper about BSS performance measures which is much newer - but it does not cover IVA settings...}
\end{comment}

%\small{
\begin{table}[htb]
\caption{Overview of various IVA methods and assumptions needed for each method.
}
\begin{center}
\resizebox{1\linewidth}{!}{
\begin{tabular}{ lccc } 
 \hline
 \textbf{Method} & \textbf{Assumptions} & \revision{\textbf{Typical applications}} & \textbf{References}\\ 
   & \textbf{in addition to (B1)-(B4)} & &\\
 \hline
  Natural gradient & The chosen distribution for sources & & \makecell{\citet{Taesu2007} \\ \citet{Anderson2012}}\\
 \hline
 FastIVA & \makecell{Nonlinearity based on\\ the choice of distribution. \\ Orthogonal unmixing matrices} & Speech source separation & \makecell{\citet{lee2007}, \\ \citet{Qian2014}} \\ 
 \hline
 AuxIVA & \makecell{Spherical symmetry assumption\\ for nonlinearity. \\ Source component vectors follow \\ super-Gaussian distribution.} & Speech source separation &\citet{Ono2011,Ono2012}\\
 \hline
 NewtonIVA & \makecell{The chosen source distribution. \\ Hessian positive definite.} & Medical signal processing &\makecell{\citet{Anderson2012}} \\ 
 \hline
 IVA-G & \makecell{Source component vectors \\ follow Gaussian distribution} & \makecell{Medical signal processing} & \makecell{\citet{Anderson2012}} \\ 
 \hline
 IVA-GGD & \makecell{Source component vectors \\ follow MGGD distribution} & Medical signal processing & \makecell{\citet{anderson2013}} \\
 \hline
 tIVA & \makecell{No further assumptions} & \makecell{Multimodal data fusion} & \citet{Adali2015} \\
 \hline
 OverIVA & \makecell{The amount of independent components \\is less than the amount of observed components. \\ 
 Source component vectors follow \\ time-varying circular Gaussian distribution or\\ a time-invariant circular Laplace distribution.\\ The separated
background noise vectors follow  \\a time-invariant Gaussian distribution.
%\\across the observed components.
\\ The sources and noise span \\orthogonal subspaces after separation.
 %See Section \ref{otherIVAmodels}
 } & Speech source separation & \makecell{\citet{Scheibler2019} \\ \citet{IkeshitaOverIVA} \\ \citet{Guo2023}} \\
 \hline
% Blinky IVA & \makecell{Sources follow time-varying \\
% spherical Gaussian distribution} & \citet{Scheibler2019blinky} \\
% \hline
  GCAV-IVA & \makecell{Constrains are non-negative. \\ Same assumptions as in AuxIVA. }& Speech source separation & \makecell{\citet{Li2020geom} \\ \citet{Goto2022}}\\
 \hline
  GC-AuxIVA-ISS & \makecell{Constrains are non-negative. \\ Same assumptions as in AuxIVA. }& Speech source separation & \makecell{\citet{Goto2022} \\ \citet{Goto2022ISS}}\\
 \hline
 FasterIVA & \makecell{Source component vectors follow \\ multivariate super-Gaussian distribution \\
  Unmixing matrices \\ constrained to be orthogonal.} & Speech source separation & \makecell{\citet{Andreas2021}} \\
 \hline
 FastDIVA & \makecell{Samples assumed to be \\  iid within blocks.\\ 
 Background probabilistic model\\ is circular Gaussian distributed. \\ CSV-separable \\ mixtures assumption.} & Speech source separation & \citet{Koldovsky2021}\\
% \hline
% \makecell{IVA-M-EMK} & \makecell{} & \makecell{\citet{Damasceno2021}} \\
 \hline
\end{tabular}}
\end{center}
\label{Table-assumptions}
\end{table}

\begin{table}[htb]
\caption{Overview of R, Matlaba and Python implementations of various IVA methods.}
\begin{center}
\begin{tabular}{lcccc} 
 \hline
 \textbf{Method} & \textbf{R} & \textbf{Matlab} & \textbf{Python} & \textbf{Reference} \\ 
 \hline
  Natural gradient &  & X & X &\makecell{https://github.com/teradepth/iva\\https://github.com/tky823/ssspy}\\
 \hline
 FastIVA & X & X & X & \makecell{\citet{Sipila2022package}\\https://github.com/teradepth/iva\\https://github.com/tky823/ssspy}\\
 \hline
  AuxIVA &  & X & X & \makecell{https://github.com/d-kitamura/AuxIVA-ISS \\ \citet{ScheiblerPython}}\\
 \hline
  NewtonIVA & X & X & X & \makecell{\citet{Sipila2022package} \\https://mlsp.umbc.edu/resources.html\\ \citet{Lehmann2022}}\\ 
 \hline
 IVA-G & X & X & X & \makecell{\citet{Sipila2022package} \\ https://mlsp.umbc.edu/resources.html \\ \citet{Lehmann2022}}\\ 
 \hline
  IVA-GGD &  & X & & https://mlsp.umbc.edu/resources.html\\
 \hline
  IVA-A-GGD &  & X & & https://mlsp.umbc.edu/resources.html\\
 \hline
  tIVA &  & X & & https://github.com/trendscenter/fit\\
 \hline
  OverIVA &  &  & X & https://github.com/onolab-tmu/overiva\\
 \hline
  GCAV-IVA &  &  &  &\\
 \hline
 FasterIVA &  &  & X & https://github.com/tky823/ssspy\\
 \hline
 FastDIVA &  & X & & https://asap.ite.tul.cz/downloads/ice/ \\
 \hline
 IVA-GL &  & X & & https://github.com/trendscenter/gift\\
 \hline
 IVA-L &  & X & & https://github.com/trendscenter/gift\\
 \hline
 Blinky IVA &  &  & X & https://github.com/onolab-tmu/blinky-iva \\
 \hline
  SparseAuxIVA & & & X & \citet{ScheiblerPython}\\
 \hline
\end{tabular}
\end{center}
\label{Table-implementations}
\end{table}

\subsection{Other IVA models}\label{otherIVAmodels}
Like in the case of ICA, various extensions of the original IVA model have been developed for different types of situations.\\

\textbf{Transposed IVA (tIVA)}: tIVA was introduced in \citet{Adali2015} to study multimodal data fusion using source separation. Multimodal data fusion is the process of combining disparate data streams (of different dimensionality, resolution, type, etc.) to generate information in a form that is more understandable or usable.
The tIVA model can be written as
\[
(\boldsymbol{x}_i^{[k]})^{\top} = (\boldsymbol{\Omega}^{[k]}\boldsymbol{s}_i^{[k]})^{\top}=(\boldsymbol{s}_i^{[k]})^{\top}(\boldsymbol{\Omega}^{[k]})^{\top}, \; \; \text{$i = 1,\dots,n$ and $k = 1,\dots,K$}.
\]
Here, the roles of the mixing matrix $\boldsymbol{\Omega}^{[k]}$ and independent sources $\boldsymbol{s}_i^{[k]}$ are reversed. In tIVA profiles, i.e., $(\boldsymbol{\Omega}^{[k]})^{\top}$ are maximally dependent across the modalities and independent among themselves within the modality. One weakness of tIVA is that it needs a significant amount of subjects to have sufficient statistical power. This is because the subjects take the place of samples after the transpose. 
Because tIVA is transposed IVA it will have the same model identification conditions as IVA.
The tIVA model generalizes the model in \citet{Correa2008}, where they used multiset canonical correlation analysis (MCCA) to maximize correlations across modalities. For the algorithm for tIVA, see \citet{Adali2015}.\\

\textbf{Overdetermined IVA (OverIVA)}: OverIVA \citep{Scheibler2019} expands IVA from the determined case to the overdetermined case. The determined case means that the amount of independent components is the same as the amount of observed components, and the overdetermined case means that the amount of independent components is less than the observed components.
The observed components $\boldsymbol{x}_i^{[k]} = (x_{i1}^{[k]},\dots, x_{im}^{[k]})$ are $m$-dimensional and the model is
\begin{align*}
    \boldsymbol{x}_{i}^{[k]} = \boldsymbol{\Omega}^{[k]}\boldsymbol{s}_i^{[k]}
    + \boldsymbol{\Psi}^{[k]}\boldsymbol{z}_i^{[k]},
\end{align*}
where $\boldsymbol{s}_i^{[k]} = (s_{i1}^{[k]}, \dots , s_{ip}^{[k]})$ are $p$-dimensional independent components, $\boldsymbol{z}_i^{[k]}$ are $(m-p)$-dimensional vector of noise, and $\boldsymbol{\Omega}^{[k]}$ and $\boldsymbol{\Psi}^{[k]}$ are the respective $m\times p$- and $m \times (m-p)$-dimensional mixing matrices. Now the $m \times m$ unmixing matrix estimate $\hat{\boldsymbol{W}}^{[k]}$ recovers the $\boldsymbol{s}_i^{[k]}$ from
\begin{align*}
    \begin{bmatrix}
        \boldsymbol{s}_i^{[k]} \\
        \boldsymbol{\Phi}^{[k]}\boldsymbol{z}_i^{[k]} \\
    \end{bmatrix}
    = \hat{\boldsymbol{W}}^{[k]}\boldsymbol{x}_i^{[k]}.
\end{align*}
The $\boldsymbol{\Phi}^{[k]}$ is an arbitrary invertible linear transformation, and the $\boldsymbol{\Phi}^{[k]}$ can be chosen to simplify the task. \citet{Scheibler2019} chose it to be
\begin{align*}
    \hat{\boldsymbol{W}}^{[k]}
    =
    \begin{bmatrix}
        \boldsymbol{W}^{[k]} \\
        \boldsymbol{U}^{[k]} \\
    \end{bmatrix},
\end{align*}
where $\boldsymbol{W}^{[k]} = (\boldsymbol{w}_1^{[k]}, \dots ,\boldsymbol{w}_p^{[k]})^\top$ is $p \times m$, $\boldsymbol{U}^{[k]} = ( \boldsymbol{J}^{[k]} \; \; -\boldsymbol{I}_{m-p})$ is $(m-p) \times m$ and $\boldsymbol{J}^{[k]}$ is $(m-p)\times p$. \citet{Scheibler2019} also assumed that the separated sources have a time-varying circular Gaussian distribution, the separated background noise vectors have a time-invariant
complex Gaussian distribution across the observed components and the sources and background span orthogonal subspaces after separation.

To solve the problem, one can use AuxIVA to minimize $\boldsymbol{W}^{[k]}$. Once AuxIVA has been applied, one must modify the lower part of the unmixing matrix, i.e., $\boldsymbol{J}^{[k]}$, so that the noise subspace remains orthogonal. For fixed $\boldsymbol{W}^{[k]}$ the $\boldsymbol{J}^{[k]}$ can be solved. For more details about the OverIVA algorithm\revision{, the complexity of the algorithm} and the form of $\boldsymbol{J}^{[k]}$, see \citet{Scheibler2019}. More about OverIVA can be found from \citet{IkeshitaOverIVA,Guo2023}.\\

\textbf{Joint IVA (jIVA)}: jIVA was introduced in \citet{GABRIELSON2020101948}, to obtain discriminating features, i.e., interpretable signatures from medical data that can be used to study differences between multiple conditions or groups. The jIVA model is a direct generalization of the joint ICA \citep{Calhoun2006jica} which will be covered in the next section. In the paper, they showed that the jIVA model is an effective method for estimating discriminating EEG features.

Let there be $K$ datasets and $T$ measurement periods. Now, the matrix form of the jIVA model is
\begin{align*}
    \bar{\boldsymbol{X}}^{[k]} = \bar{\boldsymbol{\Omega}}^{[k]}\bar{\boldsymbol{S}}^{[k]}, \; \; \text{for $k=1,\dots,K$},
\end{align*}
where $\bar{\boldsymbol{X}}^{[k]} = [\boldsymbol{X}_1^{[k]}, \dots,\boldsymbol{X}_T^{[k]}]$, $\bar{\boldsymbol{S}}^{[k]} = [\boldsymbol{S}_1^{[k]}, \dots,\boldsymbol{S}_T^{[k]}]$ are $p \times N$ matrices, and $\bar{\boldsymbol{\Omega}}^{[k]}$ is the $p \times p$ mixing matrix for the $k$th data set. Here, $N$ is the number of samples over all data sets, $\boldsymbol{X}_t^{[k]}$ indicates the observed components measured in the time period $t$ and $\boldsymbol{S}_t^{[k]}$ the independent components in the time period $t$. Both $\boldsymbol{X}_t^{[k]},\boldsymbol{S}_t^{[k]}$ are $p \times n_{t,k}$ matrices where $n_{t,k}$ tells the amount of samples in the data set. The number of samples $n_{t,k}$ may vary across the concatenated data sets $\boldsymbol{X}_t^{[k]}$ as long as the total number of samples across concatenated datasets sums up to the same number $N$ for all data matrices $\bar{\boldsymbol{X}}^{[k]}$ i.e, $\sum_{t=1}^T n_{t,k}=N$ for all $k$. \\

\textbf{Block-wise IVA}: Block-wise IVA model was introduced by \citet{Koldovský2019} and has also been used for example in \citet{Koldovsky2021} where FastDIVA algorithm was derived. In the block-wise model the samples are divided into $Q \geq 1$ time-intervals called blocks. The block-wise varying mixing model is then 
\begin{align*}%\label{block-wiseIVA}
     \boldsymbol{x}^{[k,q]} = \boldsymbol{\Omega}^{[k,q]}\boldsymbol{s}^{[k,q]},
\end{align*}
where $k=1, \dots, K$ is the data set index, $q=1, \dots, Q$ is the block index, $\boldsymbol{\Omega}^{[k,q]}$ is the $p \times p$ mixing matrix, and $\boldsymbol{s}^{[k,q]} = (s_1^{[k,q]}, \dots, s_p^{[k,q]})^\top$ are the unknown independent components. The block-wise model reduces to the basic IVA model when $Q = 1$. When $Q > 1$ IVA can be applied to each block separately. However, this approach does not guarantee the same order of components.

Rather than using IVA for each block, independent vector extraction (IVE) \citet{Koldovský2019IVE} is used with simplification, which ensures that the components in different blocks are not permuted randomly. In \citet{Koldovsky2021} they assumed constant separating vector (CSV) in each block for data sets $k=1,\dots,K$ meaning that
\begin{align*}
    \boldsymbol{w}^{[k,1]} = \boldsymbol{w}^{[k,2]} = \dots = \boldsymbol{w}^{[k,Q]} = \boldsymbol{w}^{[k]}.
\end{align*}
There is also another simplification called constant mixing vector meaning
\begin{align*}
    \boldsymbol{\omega}^{[k,1]} = \boldsymbol{\omega}^{[k,2]} = \dots = \boldsymbol{\omega}^{[k,Q]} =  \boldsymbol{\omega}^{[k]},
\end{align*}
which also guarantees the order. \citet{Koldovsky2021} expanded the CSV assumption to CSV-separable mixtures meaning that all $p$ sources to be separated obey CSV and for each $j = 1, \dots, p$ the $j$th source obeys CSV in a reduced mixture where sources $1,\dots,j-1$ have been subtracted. The FastDIVA can be used to separate $p$ sources from CSV-separable mixtures. For additional information about the FastDIVA algorithm, block-wise statistical model and the cost function for it see \citet{Koldovsky2021}.

\section{Related methods}\label{sec:related}
There are different methods that are closely related to IVA. For example, there are different group ICA models \citep{Esposito2005,Ying2008,CALHOUN2009}, canonical correlation analysis \citep[CCA,][]{Hotelling1936} and multiset canonical correlation analysis \citep[MCCA,][]{Kettering1971,Nielsin2002} and joint independent subspace analysis \citep[JISA,][]{Lahat2016JISA}. Here we explain how these methods are related to IVA and what differences these methods have compared to it. \\ 

\textbf{Group ICA}: Group ICA models try to extend ICA to multiple data sets, as IVA does. Usually, group ICA models try to solve the problem using two different approaches. The first is to use clustering techniques that combine individual independent components across the datasets \citep{Esposito2005,ONTON200699}. The second strategy is to create aggregate data that contain all observations from all datasets, directly estimate components that are consistently expressed in the population in a single set of independent components, and then back-reconstruct estimated components to the individual data. This second approach has been used, for example, to analyze fMRI data in \cite{Calhoun2001,BECKMANN2005294}. \\

\textbf{Joint ICA (\revision{jICA})}: jICA \citep{Calhoun2006jica,Adali2015} is an example of one group ICA method and it has been used for multimodal data fusion. The \revision{jICA} model assumes that 
%joint spatially
the independent sources are linearly mixed together by a shared mixing parameter. In the jICA model, the new dataset is created by concatenating each data set horizontally next to each other. For $K$ datasets the matrix form of the jICA model is
\begin{align*}
    [\boldsymbol{X}^{[1]}, \boldsymbol{X}^{[2]}, \dots, \boldsymbol{X}^{[K]}] = \boldsymbol{\Omega} [\boldsymbol{S}^{[1]},\boldsymbol{S}^{[2]}, \dots, \boldsymbol{S}^{[K]}],
\end{align*}
where $\boldsymbol{\Omega}$ is a common mixing matrix and
\begin{align*}
    \boldsymbol{X}^{[k]} = [\boldsymbol{x}_1^{[k]}, \dots, \boldsymbol{x}_p^{[k]}]^\top \; \; \text{for $k=1,2, \dots, K$.}
\end{align*}
contains the observed data for every individual. Now, one can use Infomax algorithm to this new data set to get the group ICA estimations. Then lastly, the independent components can be recovered by back-reconstruction details in \citep{Calhoun2001}.
%More about jICA one can read from (\cite{Adali2015}; \cite{Calhoun2006jica}; ).
%The dataset $\boldsymbol{X}^{[k]}$ contains the different 

One example of a method that uses clustering strategy is group ICA by self-organizing clustering (sogICA) in \citet{Esposito2005}. This method was used to fMRI group studies. In sogICA first FastICA or \revision{Infomax} algorithm is used for each dataset separately. Then the ICA estimates from each subject are organized in one single set of components, with label that preserves the link from the component to the original subject. The components are then clustered according to their mutual similarities. 

As a measure of similarity between estimated independent components they used absolute value of their mutual correlation coefficients. After calculating all the similarities, one gets a $p \times p$ similarity matrix.
%in space for source estimates or in time for the associated basis time-courses. The similar measure they defined is the weighted sum of spatial correlation coefficients and temporal correlation between independent components with coefficients $\psi$ and $(1-\psi)$ respectively, where $\psi\in [0,1]$. As example \citet{Esposito2005} generated spatial ICA estimates and, thus $\psi=1$.
The similarity matrix is then transformed to a dissimilarity matrix, which is used as a ''distance'' matrix in the original space of components and represents the input for the clustering step. One can read more about the similarity measure, dissimilarity matrix and how dissimilarity is used in the clustering from \citet{Esposito2005}.\\

\textbf{Multiset canonical correlation analysis (MCCA)}: IVA is closely related to CCA and its generalization MCCA. IVA extends these methods by incorporating higher-order statistics and to the case where the unmixing matrix is not constrained to be orthogonal. MCCA is a technique used to analyze linear relationships among more than two sets of variables. It has already been applied in various fields, including signal processing, economics, biomedical data analysis, medical imaging, meteorology, and many others. Using the IVA notation, MCCA finds a $K$ weighting vectors $\boldsymbol{w}_j^{[1]},\dots, \boldsymbol{w}_j^{[K]}$ such that the correlation within the \revision{SCV} $\hat{\boldsymbol{s}}_j = (s_j^{[1]}, \dots, s_j^{[K]})^\top$, where $s_j^{[k]} = (\boldsymbol{w}_j^{[k]})^\top \boldsymbol{x}^{[k]}$ is maximized. This can be achieved using a deflationary approach where one estimates $\boldsymbol{w}_j^{[k]}$ for $k = 1, \dots,K$ and then proceeds the estimation for $\boldsymbol{w}_{j+1}^{[k]}$ for $k = 1, \dots, K$ such that $\boldsymbol{w}_j^{[k]}$ is orthogonal to $\boldsymbol{w}_{j+1}^{[k]}$. When $K=2$, CCA is a special case of MCCA.

There are different ways to maximize all correlations/covariances between the new variables simultaneously. \citet{Kettering1971} lists five different cost functions for MCCA,  the cost function that links IVA to MCCA is the generalized variance (GENVAR) that minimizes the determinant of the \revision{SCV} correlation matrix 
\begin{align*}
    \det(\EX(\hat{\boldsymbol{s}}_j\hat{\boldsymbol{s}}_j^\top )) = \prod_{k=1}^K \lambda_{j,k}.
\end{align*} 
This cost function above is for one source component vector correlation matrix. The GENVAR cost function can be naturally extended to be the product of the covariance eigenvalues for all $p$ estimated \revision{SCV}s and the cost function is then
\begin{align}\label{eq:eig}
    \prod_{j=1}^p \prod_{k=1}^K \lambda_{j,k}.
\end{align}
Now, if one assumes multivariate Gaussian distribution for the sources and orthogonal unmixing matrices then the IVA-G cost function (\ref{IVA-Gcost}) reduces in principle to the minimization of (\ref{eq:eig}). This means that the IVA-G cost function reduces to the GENVAR cost function and gives justification that IVA is applicable to problems where MCCA has already been used. For a more comprehensive review of other cost functions, see \citet{Kettering1971} and \citet{Nielsin2002}.\\

\textbf{Joint independent subspace analysis (JISA)} : ICA and IVA models are still not enough in some situations. Either of these models can handle situations where the independent components are vectors rather than scalars. Fortunately, ICA and IVA models have been generalized to handle this problem. Independent subspace analysis \citep[ISA,][]{Theis2006} is a generalization of ICA to the problem described above, and JISA is a generalization of IVA. JISA generalization is performed by combining IVA with ISA. Next, we give a theoretical formulation of JISA model.

Let the $n$ observations of $K$ vectors $\boldsymbol{x}_i^{[k]}$, be modeled as 
\begin{align*}
    \boldsymbol{x}_i^{[k]} = \boldsymbol{\Omega}^{[k]}\boldsymbol{s}_i^{[k]} \; \; 1\leq i \leq n, \; 1\leq k \leq K,
\end{align*}
where $\boldsymbol{\Omega}^{[k]}$ are invertible $p \times p$ matrices that may differ for all $k$ and $\boldsymbol{x}_i^{[k]}$ and $\boldsymbol{s}_i^{[k]}$ are $p \times 1$ vectors. Given partition $\boldsymbol{s}_i^{[k]} = ((\boldsymbol{s}_{1i}^{[k]})^\top, \dots, (\boldsymbol{s}_{Mi}^{[k]})^\top)^\top$, where $M \leq p$, $\boldsymbol{s}_{ji}^{[k]}$ are $m_j \times 1$ vectors, $m_j \geq 1$, $\sum_{j=1}^M m_j = p$. Now the JISA model corresponds to the assumption that all the elements of the $Km_j \times 1$ vector $\boldsymbol{s}_{ji} = [(\boldsymbol{s}_{ji}^{[1]})^\top, \dots ,(\boldsymbol{s}_{ji}^{[K]})^\top]^\top$ are statistically dependent and the pairs $(\boldsymbol{s}_{ji}, \boldsymbol{s}_{li})$ are statistically independent for all $j\neq l \in \{1, \dots , M\}$ \citep{Lahat2014jisa}. If one has $\boldsymbol{m}=(m_1,\dots,m_M)^\top$ and observations $\boldsymbol{x}_i^{[k]}$, $1\leq i \leq n, 1\leq k \leq K$ then the problem of JISA is to find linear transformations $(\boldsymbol{\Omega}^{[k]})^{-1}$, such that the source vectors $\boldsymbol{s}_{1i}, \dots, \boldsymbol{s}_{Mi}$ are
as independent as possible.

The partition of $\boldsymbol{s}_i^{[k]}$ also induces partition for the mixing matrix
$\boldsymbol{\Omega}^{[k]} = [\boldsymbol{\Omega}_1^{[k]}, \dots, \boldsymbol{\Omega}_M^{[k]}]$, where $\boldsymbol{\Omega}_j^{[k]}$ is the $j$th $p \times m_j$ column-block of $\boldsymbol{\Omega}^{[k]}$. Now the multiplicative model can be rewritten as sum of $N \leq p$ components
\begin{align*}
    \boldsymbol{x}_i^{[k]} = \sum_{j=1}^M \boldsymbol{x}_{ji}^{[k]}, \; \; \text{where $\boldsymbol{x}_{ji}^{[k]}=\boldsymbol{\Omega}_j^{[k]}\boldsymbol{s}_{ji}^{[k]}$}.
\end{align*}
Like in IVA the JISA model is not uniquely defined. Let $\boldsymbol{B}^{[k]}$ be any invertible $m_j \times m_j$ matrix then
\begin{align*}
    \boldsymbol{x}_{ji}^{[k]} = \boldsymbol{\Omega}_j^{[k]}\boldsymbol{s}_{ji}^{[k]} =
    \boldsymbol{\Omega}_j^{[k]}\boldsymbol{B}_j^{[k]}(\boldsymbol{B}_j^{[k]})^{-1}\boldsymbol{s}_{ji}^{[k]}.
\end{align*}
This means that JISA can only blindly identify the column-space of $\boldsymbol{\Omega}_j^{[k]}$\citep{Lahat2014jisa,Lahat2016JISA}. Thus, JISA is a subspace estimation problem.
 The identifiability of this model has been proved in \citet{Lahat2019JISAidentifiability} under necessary and sufficient conditions.

JISA model includes other well-known models such as IVA, ISA and ICA as special cases. If $K = 1$, then the model reduces to the ISA model. The ICA model is the special case of the ISA model, i.e., $m_j=1$ for all $j$. When $m_j=1$ for all $j$ and $K \geq 2$ the IVA model is obtained. So, one can think of the JISA model as a multiple ISA problem linked together by statistical dependencies of the latent components.

\section{Examples}\label{sec:application}

\subsection{Illustration to mixed images}\label{Ex:Image}
Next we give an illustration of IVA by separating mixed colored images. Colored images are often represented as RGB images, which consist of three different color channels, red (R), green (G) and blue (B). Now each pixel of the picture gets a value between 0 and 255 for each channel. Higher value means brighter color for the pixel.

\begin{figure}[tb]
    \centering
    \includegraphics[width=1\linewidth]{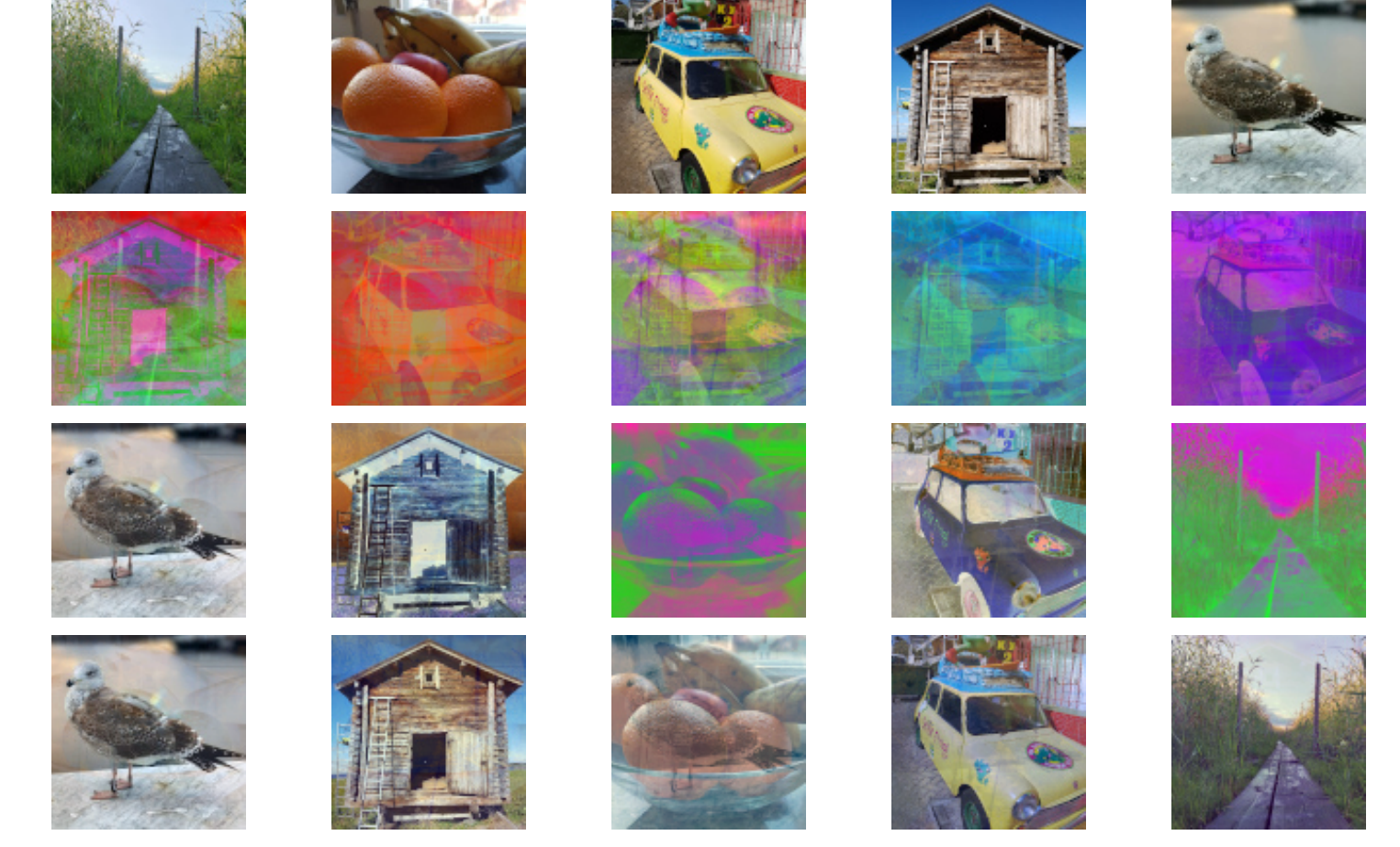}
    \caption{The original source images are presented in the first row and the mixtures of the
source images in the second row. The separated images received from IVA-G-N are in the third row and the manually corrected images in the last row. The original pictures are owned by Mika Sipilä.}
    \label{IVAEXAMPLE}
\end{figure}

In this example, we will use Newton update IVA with Gaussian source distributions to separate five original images from their mixture. The source signals are composed of $p=5$ independent $100 \times 100$ pixel images. Each row of the images is then stacked in a vector making the sample size $n = 10000$. Now, the number of datasets equals $K = 3$, where each data set corresponds to one of the RGB colors. The mixed images are then obtained by generating three different mixing matrices $\boldsymbol{\Omega}^{[1]},\boldsymbol{\Omega}^{[2]}$ and $\boldsymbol{\Omega}^{[3]}$ whose elements follow the normal distributions. In the first row of Figure \ref{IVAEXAMPLE} one can see the original pictures and in the second row the mixed images.

The estimated images can be seen in the third row of Figure \ref{IVAEXAMPLE}. Now the estimated images correspond to original images up to a sign change and permutation. The corrected pictures can be seen in the last row of Figure \ref{IVAEXAMPLE}. The results look reasonably good taking into account that the Newton update IVA is suboptimal for the image separation task. This statement can also be seen by the $ISI_{joint}$ value which was approximately \revision{$0.57$}. This tells us that the results are nowhere near optimal. One reason for suboptimality is that the Newton update IVA does not take serial dependence into account, which is why the spatial correlation structures inherent in images are ignored. \revision{The R-code for the example can be found in \url{https://github.com/miro-arvila/IVA_review}.}

\subsection{\revision{Comparison of IVA with individual ICAs for multisubject fMRI data}}\label{Ex:fMRI}

\revision{
Lastly, we illustrate the differences between several ICA analyses and an IVA analysis using
a multisubject fMRI dataset from a visuomotor task. For the fMRI analysis we use Group
ICA/IVA fMRI Toolbox (GIFT) \citep{CorreaGIFT}, which has been developed and implemented in MATLAB. GIFT provides analysis functionality, and both the software and the data can be obtained from the GIFT repository. 
%The data are available in the Group ICA/IVA fMRI Toolbox (GIFT) \citep{CorreaGIFT}, which also provides the analysis functionality and can be obtained from the GIFT repository. 
The dataset consists of preprocessed fMRI data
from three subjects, each with 220 time points collected while performing visually cued
button‐press responses; see \cite{Calhoun2001} for a detailed description of the task.
For demonstration purposes we keep the analysis simple; more sophisticated approaches are
described in the GIFT manual.
}

\revision{
As the IVA method we use IVA‐GL (described in Section~\ref{sec:est}), and for comparison we use
the Infomax ICA algorithm. Consistent with standard fMRI preprocessing, we apply both
methods to the first 20 principal components of the data and employ the default parameter
settings suggested in GIFT.
}

\revision{
The main objective for this dataset is to identify components associated with task‐related
visual processing. Based on prior knowledge, two components are expected to be strongly
task‐related and one component transiently related, with activation located in the primary
visual cortex at the posterior of the brain. Each method estimates 20 components for each
subject; however, IVA provides a consistent ordering of the components across subjects,
whereas ICA does not. Thus, to ensure comparability, we order components according to
their temporal association with the task design using a regression‐based criterion and sort
by the resulting $R^2$ values. Focusing on the highest task‐related components, IVA returns
a consistent set of components across subjects, whereas ICA yields subject‐specific orderings,
making the identification of task‐related components more laborious when analyzed
individually.
}

\begin{figure}
    \centering
    \includegraphics[width=1.05\linewidth]{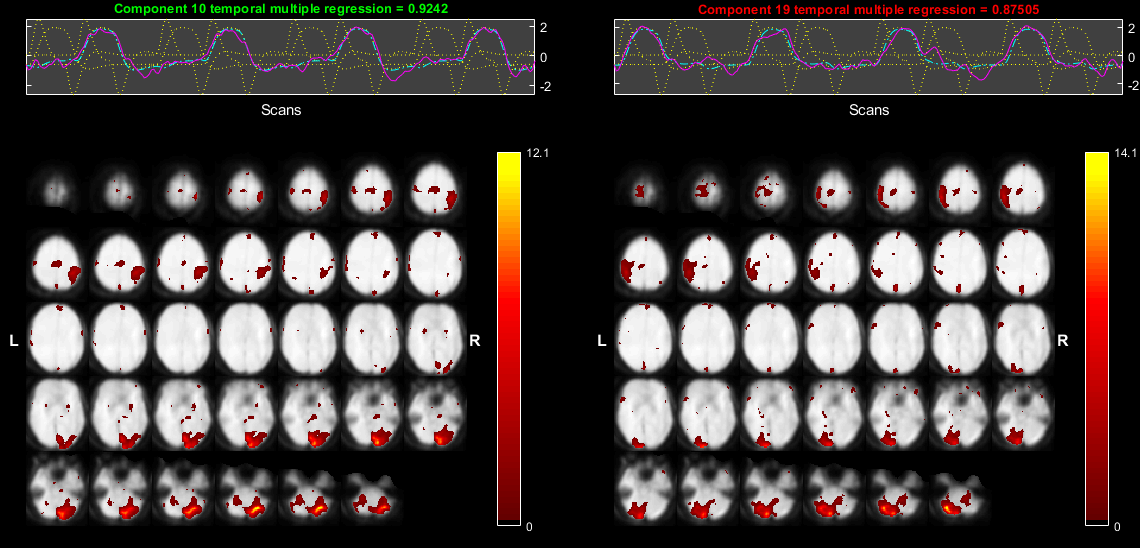}
    \includegraphics[width=1.05\linewidth]{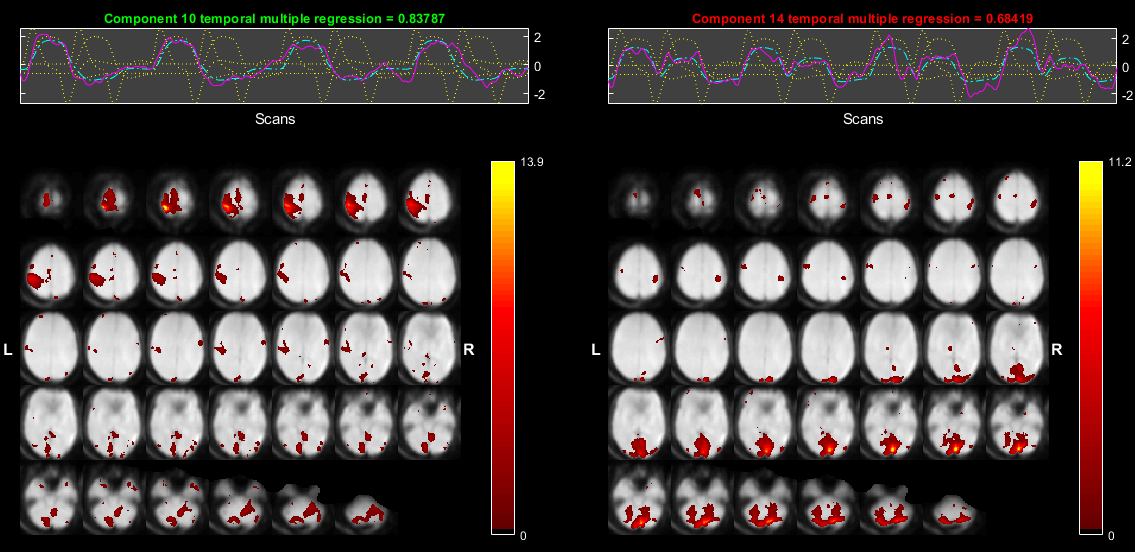}
    \includegraphics[width=1.05\linewidth]{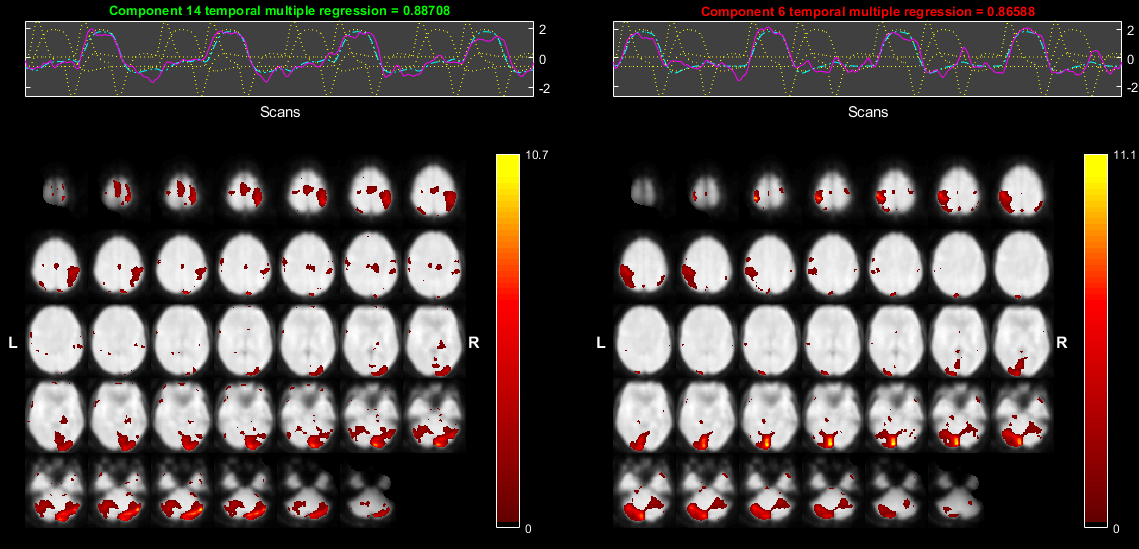}
    \caption{Rows correspond to subjects 1, 2 and 3 respectively from top to bottom and the columns present the two revealed task related components. In one row there are the two task related components, task related timecourses and the brain activation when used ICA individually to every subject. In the timecourses the purple lines are the ICA timecourses and blue lines are linefits to the ICA timecourses. The activation pictures are slices of brain from different heights. It starts at the top of the brain in the first row and goes lower with each picture.}
    \label{ICA:temporalsubj}
\end{figure}

\begin{figure}
    \centering
    \includegraphics[width=1.05\linewidth]{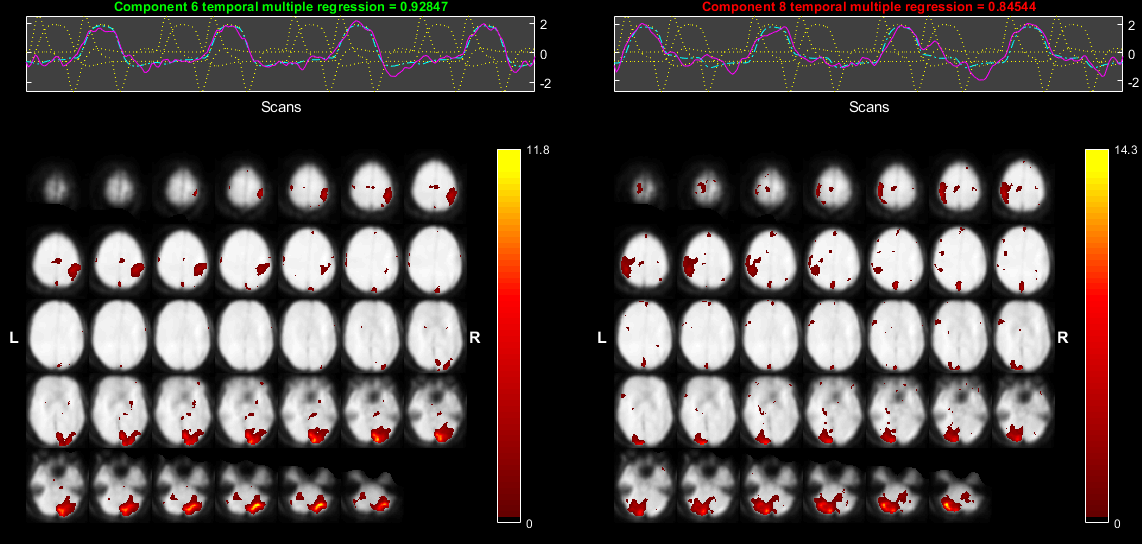}
    \includegraphics[width=1.05\linewidth]{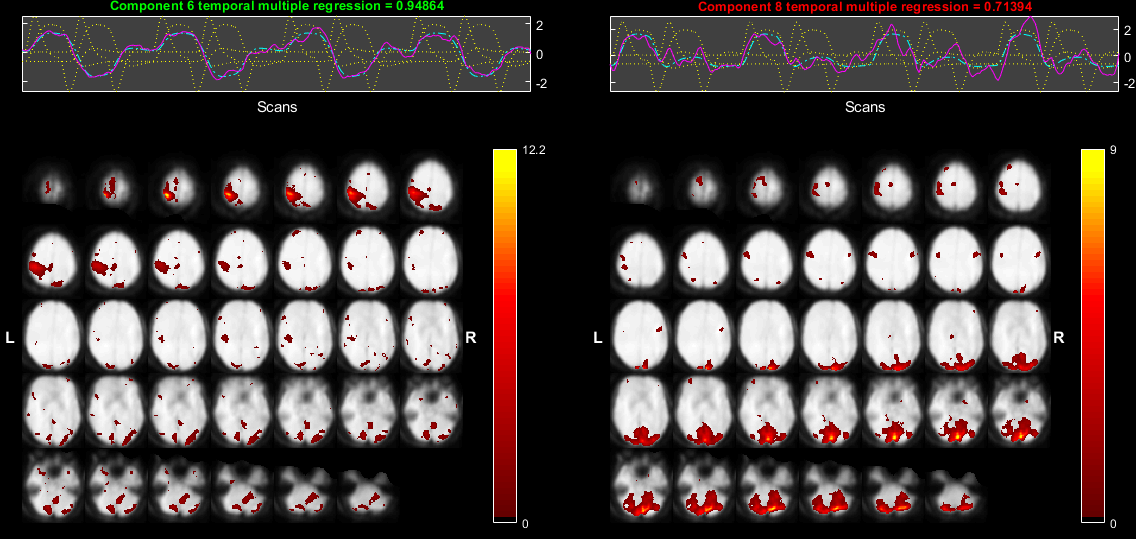}
    \includegraphics[width=1.05\linewidth]{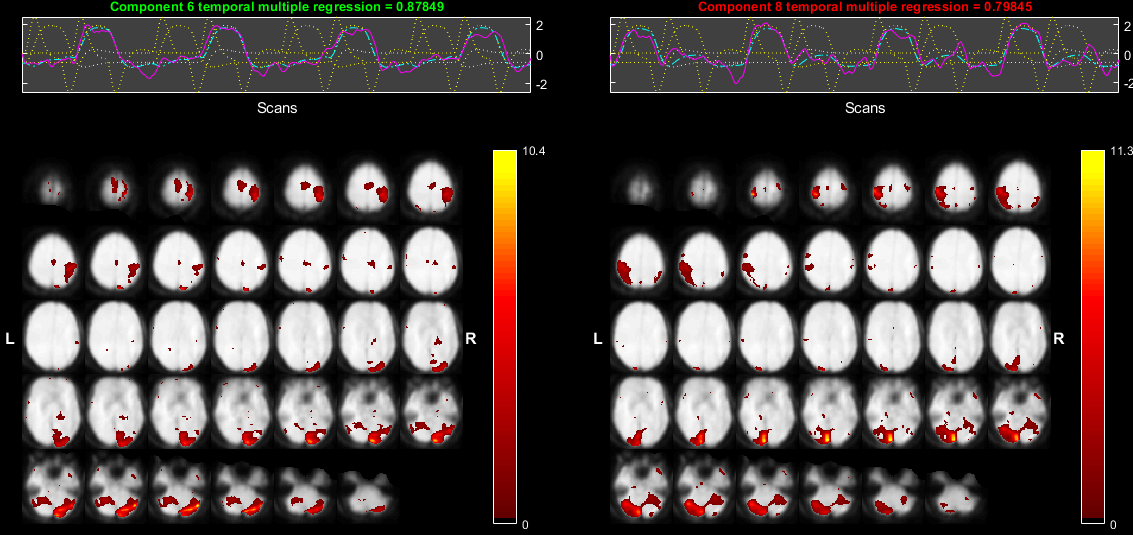}
    \caption{Rows correspond to subjects 1, 2 and 3 respectively from top to bottom and the columns present the two revealed task related components. The two task related components, task related timecourses and the brain activation when used IVA. In the timecourses the purple lines are the IVA timecourses and blue lines are linefits to the IVA timecourses. The activation pictures are slices of brain from different heights. It starts at the top of the brain in the first row and goes lower with each picture.}
    \label{IVA:temporalsubj}
\end{figure}

\revision{
The task‐related components for all subjects are visualized in
Figure~\ref{ICA:temporalsubj} (ICA) and Figure~\ref{IVA:temporalsubj} (IVA).
It is evident that both approaches detect activation in similar brain regions. In particular,
the first component predominantly shows activation in the right visual cortex, whereas the
second component shows activation in the left visual cortex. Subjects~1 and~3 exhibit similar
patterns for both methods; however, the IVA results display more focused activation within
the expected cortical areas. For Subject~2, both ICA and IVA identify left‐ and right‐hemisphere
activation but tend to combine them to varying degrees. The IVA component nevertheless
shows stronger and more spatially concentrated activation. Given the hemispheric
structure of visual processing, a separation between left and right visual cortex activation is desirable, and in this regard IVA provides a slightly improved interpretation.
}

\revision{
Based on this example, we conclude that joint estimation in IVA better exploits the
shared task structure across subjects. In addition, IVA simplifies downstream analysis,
as the component ordering is consistent and only a single permutation must be resolved.
In contrast, performing individual ICAs requires separate permutation alignment for each
subject, which becomes increasingly cumbersome in larger studies. For reproducibility and
additional details of the settings used here, as well as the full GIFT results for this example, we provide further information at \url{https://github.com/miro-arvila/IVA_review}.
}

\revision{Note that fMRI represents just one complex application area in which IVA has been used extensively.
Another closely related domain is the analysis of EEG data using IVA methods.
Interested readers may refer to, for example, \citet{Chen2014,GABRIELSON2020101948,Liu2021}.}

\section{Discussion \revision{and open problems}}\label{sec:discussion}
In this paper, we provided a comprehensive introduction to the IVA model, various methods that solve the IVA problem, and closely related approaches, highlighting their differences. We also illustrated these concepts using mixed colored images \revision{and visuomotor fMRI data}. This review shows that IVA has been and continues to be a very active area of research. Numerous methods have been proposed over the past two decades, primarily within the signal processing literature. As stated in the introduction, the main goal of this paper was to make IVA more known to statisticians.

This review focused on IVA assuming independent and identically distributed samples and real-valued source signals. IVA in case of non-iid samples has been studied, for example, in \citet{anderson2013}. Complex-valued IVA has been studied in signal processing literature, for example, in \cite{ANDERSON2012COMPLEXIVA,Mowakeaa2020,Kautský2020}.

\revision{There are still many open questions about IVA from statistical perspective}. Although various IVA methods have been applied in numerous contexts such as EEG and fMRI data, their statistical properties remain largely unexplored. Often algorithms are verified with simulation experiments, but the statistical properties such as asymptotical distributions of the methods are not studied. As an example, the asymptotical properties of FastIVA have not been studied in the same way as asymptotical properties of FastICA \citep{Ollila:2010,Nordhausenetal:2011,MiettinenNordhausenOjaTaskinen2014,Miettinenetal:2017}. 

\revision{Bayesian framework in IVA is also largely unexplored. Bayesian models in IVA context have been studied in \citet{Brendel2020EVD,Brendel2020spat,Brendel2020prob} but this could be explored more. Another interesting direction would be to explore more semi-parametric methods for IVA. For example \citet{Fu2015} solved IVA problem via minimizing entropy rate bound where the estimation of entropy was done by semi-parametric approach. \citet{Damasceno2021} developed a multivariate density estimation technique for IVA problem which was based on semi-parametric method called maximum entropy principle.}

\revision{Other interesting research questions would include under- and overdetermined IVA models. In underdetermined case the amount of independent components is larger than the amount of observed components. Overdetermined IVA has been studied for example in \citet{Scheibler2019,IkeshitaOverIVA,Guo2023} but underdetermined case still seems to be unexplored. From the side of ICA the underdetermined case has been studied, for example, in \citet{Araki2004,Kim2004,ZHANG20061538}.}

Another aspect that has been studied in the ICA context but not extensively in the IVA context is the analysis that involves different data types. \revision{In future it would be interesting to extend IVA to different data types, for example, to tensor and functional data which  \citet{virta2017independent,VirtaLiNordhausenOja2020} did in ICA case.} As highlighted in this review and the preceding discussion, the contribution of statisticians in this area would provide significant value.

\section*{Acknowledgments}
We acknowledge the support from the Research Council of Finland (356484) to MA and ST, the support from the Research Council of Finland (363261) to KN, the support from Vilho, Yrjö and Kalle Väisälä foundation to MS, and the support from the HiTEc COST Action (CA21163) to KN and ST.

\begin{appendix}
    \section{Appendix}\label{sec:appendix}

\subsection{ICA structure}\label{ICA figure}
\begin{figure}[h]
    \centering
    \includegraphics[width=1\linewidth]{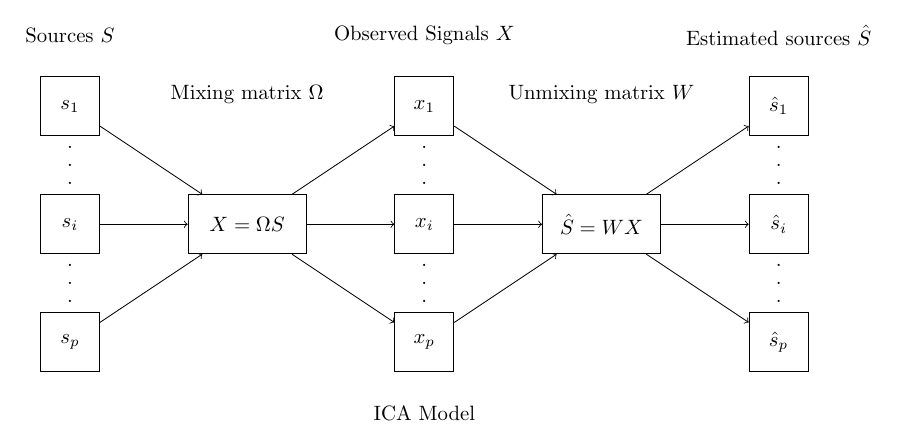}
    \caption{Structure of ICA model with $p$ sources and $p$ observed signals. The graph illustrates the mixing process from independent source components to observations and the unmixing process from observations to source estimates.} 
    \label{ICAMODELpic}
\end{figure}

\revision{Figure \ref{ICAMODELpic} shows a schematic representation of the ICA model.}

\subsection{Information theory} \label{information theory}
Mutual information $\mathcal{I}(\cdot)$ is a measure of the information that members of a set of random variables have on the other random variables in the set. Statistically this means that mutual information tells something about the statistical independence of the random variables. In ICA, mutual information can be used as a cost function. Mutual information $\mathcal{I}(\cdot)$ for $p$-variate random vector $\boldsymbol{y}$ can be defined using the entropy or the Kullback-Leibler divergence \citep{Kullback1951}. Here we give the definition using the Kullback-Leibler divergence, i.e., 
\[
\mathcal{I}(\boldsymbol{y}) = D_{KL}(P(\boldsymbol{y})|Q(\boldsymbol{y})),
\]
where Kullback-Leibler divergence $D_{KL}(\cdot|\cdot)$ is defined between two $n$-dimensional probability density functions $P,Q$ as
\begin{equation*}
    D_{KL}(P|Q) = \int P(\boldsymbol{y})\log\left(\frac{P(\boldsymbol{y})}{Q(\boldsymbol{y})}\right)d\boldsymbol{y}.
\end{equation*}
%$P$ and $Q$ are probability distributions for the variable $\boldsymbol{y}$.

Differential entropy for continuous random vector is defined as 
\[
\mathcal{H}(\boldsymbol{y}) = \EX(-\log(P(\boldsymbol{y}))) = -\int P(\boldsymbol{y})\log(P(\boldsymbol{y}))d\boldsymbol{y},
\]
and differential cross entropy is defined as
\[
\mathcal{H}(\boldsymbol{y}:P,Q) = -\EX_P(\log(Q(\boldsymbol{y}))) = - \int P(\boldsymbol{y})\log(Q(\boldsymbol{y}))d\boldsymbol{y},
\]
where $P$ and $Q$ are probability distributions. Here $\EX_P(\cdot)$ means taking the expected value operator with respect to the distribution $P$. Note that entropy has a transformation property meaning that if $\boldsymbol{v}=f(\boldsymbol{y})$ and that the transformation is invertible, then $\mathcal{H}(\boldsymbol{v}) = \mathcal{H}(\boldsymbol{y}) + \EX(\log|\det Jf(\boldsymbol{y})|)$, where $Jf(\boldsymbol{y})$ is the Jacobian matrix of the function $f$ evaluated at the point $\boldsymbol{y}$. 

One can write Kullback-Leibler divergence using differential entropy and differential cross entropy
\begin{align*}
    D_{KL}(P|Q) &= \int P(\boldsymbol{y})\log\left(\frac{P(\boldsymbol{y})}{Q(\boldsymbol{y})}\right)d\boldsymbol{y} \\
    &= \int P(\boldsymbol{y})\log\left(P(\boldsymbol{y})\right)d\boldsymbol{y}
    - \int P(\boldsymbol{y})\log\left(Q(\boldsymbol{y})\right)d\boldsymbol{y} \\
    &= \mathcal{H}(\boldsymbol{y}:P,Q)-\mathcal{H}(\boldsymbol{y}).
\end{align*}

\subsection{Negentropy} \label{negen}

Negentropy for random vector $\boldsymbol{x}=(x_1,\dots,x_p)^\top$ is defined as
\[
\mathcal{J}(\boldsymbol{x}) = \mathcal{H}(\boldsymbol{x}_{gauss}) - \mathcal{H}(\boldsymbol{x}),
\]
where $\mathcal{H}(\boldsymbol{x}_{gauss})$ is a Gaussian random vector of the same covariance (or correlation)
matrix as $\boldsymbol{x}$. The entropy for $\boldsymbol{x}_{gauss}$ can be calculated and it is
\begin{align*}
    \mathcal{H}(\boldsymbol{x}_{gauss}) = \frac{1}{2}\log(\det(\boldsymbol{\Sigma})) + \frac{p}{2}(1 + \log(2\pi)),
\end{align*}
where $\boldsymbol{\Sigma}$ is the covariance matrix of $\boldsymbol{x}$.

Let $\boldsymbol{y}=(y_1,\dots,y_p)^\top$ be a random vector such that $\boldsymbol{y}=\boldsymbol{Wx}$. Then it can be shown that
\begin{align*}
    \mathcal{J}(\boldsymbol{y}) = \mathcal{J}(\boldsymbol{x})
\end{align*}
for invertible $\boldsymbol{W}$. This means that negentropy is invariant for invertible linear transformations \citep{Comon1994}.

Also for uncorrelated $y_j$ with unit variance one can show that
\[
\mathcal{I}(\boldsymbol{y}) = \alpha - \sum_{j=1}^p \mathcal{J}(y_j),
\]
where $\alpha$ is a constant which is not dependent on unmixing matrix $\boldsymbol{W}$.

Next we give two approximations of negentropy, which has been used in the ICA context. One of the classical methods to estimate negentropy of random variable $y$ is to use the higher order moments as in \citet{Jones1987}:
\begin{align*}
    \mathcal{J}(y) \approx \frac{1}{12}\EX(y^3)^2 + \frac{1}{48}kurt(y)^2,
\end{align*}
where $kurt(y) = \EX(y^4) - 3(\EX(y^2))^2$ is the kurtosis of the random variable $y$. The random variable $y$ is assumed here to have zero mean and unit variance, which reduces $kurt(y) = \EX(y^4) -3$. This kind of approximation suffers from nonrobustness because kurtosis is very sensitive to outlying observations.

Another way to approximate negentropy which does not suffer from the same problem as kurtosis, is to use expectations of nonquadratic functions $G$ \citep{Hyvarinen1997negappox}. These approximations were based on maximum-entropy principle and are the form of
\begin{align*}
    \mathcal{J}(y) \approx \sum_{l = 1}^m k_l(\EX(G_l(y)) - \EX(G_l(\nu)))^2,
\end{align*}
where $k_l$ are positive constants, $\nu$ is a standardized Gaussian random variable. If only one nonquadratic function $G$ is used then the approximation is
\begin{align*}
    \mathcal{J}(y) \approx (\EX(G(y)) - \EX(G(\nu)))^2.
\end{align*}
The approximation with nonquadratic functions is a generalization of the approximation with higher order moments. One can see this by approximating negentropy with two function $G_1$ and $G_2$ such that $G_1$ is odd and $G_2$ is even. Then the approximation is 
\begin{align*}
    \mathcal{J}(y) \approx k_1(\EX(G_1(y)))^2 + k_2(\EX(G_2(y)) - \EX(G_2(\nu)))^2,
\end{align*}
where $k_1$ and $k_2$ are positive constants. When $G_1(y) = y^3$ and $G_2(y)=y^4$, the approximation reduces to the moment approximation. For more information about the approximations, see \citet{HYVARINEN2000411}.

\subsection{Distributions used in FastIVA}\label{FastIVAapp}
%\textcolor{red}{Why not give these in distributions section. Why the covariance is ignored here?}

\citet{lee2007} used two probability density functions in the FastIVA and one entropy approximation. The first pdf which they introduced was a spherically symmetric, exponential norm distribution (SEND),
\begin{align*}
    p(\hat{\boldsymbol{s}}_j) \propto \frac{\exp(-\sqrt{(4/K)}\left\Vert \hat{\boldsymbol{s}}_j \right\Vert_2)}
    {\left\Vert \hat{\boldsymbol{s}}_j \right\Vert_2^{K-1}},
\end{align*}
where $K$ is the dimension of real domain. The second pdf was spherically symmetric Laplace distribution (SSL), which had been proposed by \citet{Hyvarinen2000ISA} to ISA. The pdf of SSL is 
\begin{align*}
     p(\hat{\boldsymbol{s}}_j) \propto \exp(-\sqrt{2(K+1)}\left\Vert \hat{\boldsymbol{s}}_j \right\Vert_2).
\end{align*}

\begin{comment}
\subsection{Individual subject pictures from visuomotor data}\label{fMRI_subject_pictures}
\revision{In the Figure \ref{IVA:subjects} one can see the brain activation and the time series of the components for all three subjects using IVA and in Figure \ref{GICA:subjects} using GICA.}

    \begin{figure}
    \centering
    \includegraphics[width=0.75\linewidth]{mreg_spatsort_IVA_subj1.png}
    \includegraphics[width=0.75\linewidth]{mreg_spatsort_IVA_subj2.png}
    \includegraphics[width=0.75\linewidth]{mreg_spatsort_IVA_subj3.png}
    \caption{The brain activation areas and the time series for the two components using IVA for all three different subjects}
    \label{IVA:subjects}
\end{figure}

\begin{figure}
    \centering
    \includegraphics[width=0.75\linewidth]{mreg_spatsort_GICA_subj1.png}
    \includegraphics[width=0.75\linewidth]{mreg_spatsort_GICA_subj2.png}
    \includegraphics[width=0.75\linewidth]{mreg_spatsort_GICA_subj3.png}
    \caption{The brain activation areas and the time series for the two components using GICA for all three different subjects}
    \label{GICA:subjects}
\end{figure}
\end{comment}

\end{appendix}

\bibliographystyle{imsart-nameyear} % Style BST file (imsart-number.bst or imsart-nameyear.bst)
\bibliography{library}  

@INPROCEEDINGS{CorreaGIFT,
  author={Correa, N. and Adali, T. and Yi-Ou Li and Calhoun, V.D.},
  booktitle={Proceedings. (ICASSP '05). IEEE International Conference on Acoustics, Speech, and Signal Processing, 2005.}, 
  title={Comparison of blind source separation algorithms for FMRI using a new Matlab toolbox: GIFT}, 
  year={2005},
  volume={5},
  number={},
  pages={v/401-v/404 Vol. 5},
  doi={10.1109/ICASSP.2005.1416325}
}

@InProceedings{Kim2004,
author="Kim, Sang Gyun
and Yoo, Chang D.",
editor="Puntonet, Carlos G.
and Prieto, Alberto",
title="Underdetermined Independent Component Analysis by Data Generation",
booktitle="Independent Component Analysis and Blind Signal Separation",
year="2004",
publisher="Springer Berlin Heidelberg",
address="Berlin, Heidelberg",
pages="445--452"
}

@article{ZHANG20061538,
title = {A Gaussian mixture model for underdetermined independent component analysis},
journal = {Signal Processing},
volume = {86},
number = {7},
pages = {1538-1549},
year = {2006},
issn = {0165-1684},
doi = {https://doi.org/10.1016/j.sigpro.2005.08.009},
url = {https://www.sciencedirect.com/science/article/pii/S0165168405002732},
author = {Yingyu Zhang and Xizhi Shi and Chi Hau Chen}
}

@INPROCEEDINGS{Araki2004,
  author={Araki, S. and Makino, S. and Blin, A. and Mukai, R. and Sawada, H.},
  booktitle={2004 IEEE International Conference on Acoustics, Speech, and Signal Processing}, 
  title={Underdetermined blind separation for speech in real environments with sparseness and ICA}, 
  year={2004},
  volume={3},
  number={},
  pages={iii-881},
  doi={10.1109/ICASSP.2004.1326686}
}

@ARTICLE{Liu2021,
  author={Liu, Aiping and Song, Gongzheng and Lee, Soojin and Fu, Xueyang and Chen, Xun},
  journal={IEEE Transactions on Instrumentation and Measurement}, 
  title={A State-Dependent IVA Model for Muscle Artifacts Removal From EEG Recordings}, 
  year={2021},
  volume={70},
  number={},
  pages={1-13},
  doi={10.1109/TIM.2021.3071217}
}

@ARTICLE{Chen2014,
  author={Chen, Xun and Liu, Aiping and McKeown, Martin J. and Poizner, Howard and Wang, Z. Jane},
  journal={IEEE Transactions on Biomedical Engineering}, 
  title={An EEMD-IVA Framework for Concurrent Multidimensional EEG and Unidimensional Kinematic Data Analysis}, 
  year={2014},
  volume={61},
  number={7},
  pages={2187-2198},
  doi={10.1109/TBME.2014.2319294}
}

@INPROCEEDINGS{Engberg2016,
  author={E. Engberg, Astrid M. and Andersen, Kasper W. and Mørup, Morten and Madsen, Kristoffer H.},
  booktitle={2016 International Workshop on Pattern Recognition in Neuroimaging (PRNI)}, 
  title={Independent vector analysis for capturing common components in fMRI group analysis}, 
  year={2016},
  volume={},
  number={},
  pages={1-4},
  doi={10.1109/PRNI.2016.7552351}
}

@INPROCEEDINGS{Goto2022ISS,
  author={Goto, Kana and Ueda, Tetsuya and Li, Li and Yamada, Takeshi and Makino, Shoji},
  booktitle={2022 30th European Signal Processing Conference (EUSIPCO)}, 
  title={Geometrically Constrained Independent Vector Analysis with Auxiliary Function Approach and Iterative Source Steering}, 
  year={2022},
  volume={},
  number={},
  pages={757-761},
  doi={10.23919/EUSIPCO55093.2022.9909912}
}

@article{Wang2025,
author = {Wang, Dahu and Liu, Chang},
title = {An Improved Fast Independent Vector Analysis Algorithm for Enhanced Efficiency},
journal = {Electronics Letters},
volume = {61},
number = {1},
pages = {e70246},
doi = {https://doi.org/10.1049/ell2.70246},
url = {https://ietresearch.onlinelibrary.wiley.com/doi/abs/10.1049/ell2.70246},
eprint = {https://ietresearch.onlinelibrary.wiley.com/doi/pdf/10.1049/ell2.70246},
year = {2025}
}

@INPROCEEDINGS{Fu2015,
  author={Fu, Geng-Shen and Anderson, Matthew and Adalı, Tülay},
  booktitle={2015 49th Annual Conference on Information Sciences and Systems (CISS)}, 
  title={Independent vector analysis by entropy rate bound minimization}, 
  year={2015},
  volume={},
  number={},
  pages={1-6},
  doi={10.1109/CISS.2015.7086842}
}

@ARTICLE{Brendel2020prob,
  author={Brendel, Andreas and Haubner, Thomas and Kellermann, Walter},
  journal={IEEE Transactions on Signal Processing}, 
  title={A Unified Probabilistic View on Spatially Informed Source Separation and Extraction Based on Independent Vector Analysis}, 
  year={2020},
  volume={68},
  number={},
  pages={3545-3558},
  doi={10.1109/TSP.2020.3000199}
}

@INPROCEEDINGS{Brendel2020EVD,
  author={Brendel, Andreas and Kellermann, Walter},
  booktitle={2020 28th European Signal Processing Conference (EUSIPCO)}, 
  title={Informed Source Extraction based on Independent Vector Analysis using Eigenvalue Decomposition}, 
  year={2021},
  volume={},
  number={},
  pages={875-879},
  doi={10.23919/Eusipco47968.2020.9287733}
}

@ARTICLE{Koldovský2019IVE,
  author={Koldovský, Zbyněk and Tichavský, Petr},
  journal={IEEE Transactions on Signal Processing}, 
  title={Gradient Algorithms for Complex Non-Gaussian Independent Component/Vector Extraction, Question of Convergence}, 
  year={2019},
  volume={67},
  number={4},
  pages={1050-1064},
  doi={10.1109/TSP.2018.2887185}
}

@INPROCEEDINGS{Koldovský2019,
  author={Koldovský, Zbyněk and Málek, Jiří and Janský, Jakub},
  booktitle={ICASSP 2019 - 2019 IEEE International Conference on Acoustics, Speech and Signal Processing (ICASSP)}, 
  title={Extraction of Independent Vector Component from Underdetermined Mixtures through Block-wise Determined Modeling}, 
  year={2019},
  volume={},
  number={},
  pages={7903-7907},
  doi={10.1109/ICASSP.2019.8683431}
}

@ARTICLE{Taesu2007,
  author={Kim, Taesu and Attias, Hagai T. and Lee, Soo-Young and Lee, Te-Won},
  journal={IEEE Transactions on Audio, Speech, and Language Processing}, 
  title={Blind Source Separation Exploiting Higher-Order Frequency Dependencies}, 
  year={2007},
  volume={15},
  number={1},
  pages={70-79},
  doi={10.1109/TASL.2006.872618}
}

@ARTICLE{Lee2007entropicapprox,
  author={Lee, Intae and Lee, Te-Won},
  journal={IEEE Transactions on Audio, Speech, and Language Processing}, 
  title={On the Assumption of Spherical Symmetry and Sparseness for the Frequency-Domain Speech Model}, 
  year={2007},
  volume={15},
  number={5},
  pages={1521-1528},
  doi={10.1109/TASL.2007.899231}
}

@article{Chang01062008,
author = {Ann‐Chen Chang and Chih‐Wei Jen and Ing‐Jiunn Su},
title = {2‐D DOA estimation for local scattered CDMA signals by modified Fastica in colored non‐Gaussian noise},
journal = {Journal of the Chinese Institute of Engineers},
volume = {31},
number = {4},
pages = {691--696},
year = {2008},
publisher = {Taylor \& Francis},
doi = {10.1080/02533839.2008.9671421},
URL = {https://doi.org/10.1080/02533839.2008.9671421},
eprint = {https://doi.org/10.1080/02533839.2008.9671421}
}

@ARTICLE{Prasad2005,
author={Prasad, Rajkishore  and Saruwatari, Hiroshi and Shikano, Kiyohiro},
journal={IEICE TRANSACTIONS on Fundamentals},
title={Blind Separation of Speech by Fixed-Point ICA with Source Adaptive Negentropy Approximation},
year={2005},
volume={E88-A},
number={7},
pages={1683-1692},
doi={10.1093/ietfec/e88-a.7.1683},
ISSN={},
month={July}
}

@article{ANDERSON2012COMPLEXIVA,
title = {Complex-valued independent vector analysis: Application to multivariate Gaussian model},
journal = {Signal Processing},
volume = {92},
number = {8},
pages = {1821-1831},
year = {2012},
note = {Latent Variable Analysis and Signal Separation},
issn = {0165-1684},
doi = {https://doi.org/10.1016/j.sigpro.2011.09.034},
url = {https://www.sciencedirect.com/science/article/pii/S0165168411003549},
author = {Matthew Anderson and Xi-Lin Li and Tülay Adalı}
}

@ARTICLE{Kautský2020,
  author={Kautský, Václav and Tichavský, Petr and Koldovský, Zbyněk and Adalı, Tülay},
  journal={IEEE Transactions on Signal Processing}, 
  title={Performance Bounds for Complex-Valued Independent Vector Analysis}, 
  year={2020},
  volume={68},
  number={},
  pages={4258-4267},
  doi={10.1109/TSP.2020.3009507}
}

@manual{Sipila2022package,
    title = {ivaBSS: Tools for Independent Vector Analysis},
    author = {Mika Sipilä and Klaus Nordhausen and Sara Taskinen},
    year = {2022},
    note = {R package version 1.0.0},
    url = {https://CRAN.R-project.org/package=ivaBSS}
}

@INPROCEEDINGS{Lehmann2022,
  author={Lehmann, Isabell and Acar, Evrim and Hasija, Tanuj and Akhonda, M.A.B.S. and Calhoun, Vince D. and Schreier, Peter J. and Adali, Tülay},
  booktitle={ICASSP 2022 - 2022 IEEE International Conference on Acoustics, Speech and Signal Processing (ICASSP)}, 
  title={Multi-Task fMRI Data Fusion Using IVA and PARAFAC2}, 
  year={2022},
  volume={},
  number={},
  pages={1466-1470},
  doi={10.1109/ICASSP43922.2022.9747662}
}

@INPROCEEDINGS{ScheiblerPython,
  author={Scheibler, Robin and Bezzam, Eric and Dokmanić, Ivan},
  booktitle={2018 IEEE International Conference on Acoustics, Speech and Signal Processing (ICASSP)}, 
  title={Pyroomacoustics: A Python package for audio room simulation and array processing algorithms}, 
  year={2018},
  volume={},
  number={},
  pages={351-355},
  doi={10.1109/ICASSP.2018.8461310}
}

@INPROCEEDINGS{Janský2016,
  author={Janský, Jakub and Koldovský, Zbyněk and Ono, Nobutaka},
  booktitle={2016 IEEE International Workshop on Acoustic Signal Enhancement (IWAENC)}, 
  title={A computationally cheaper method for blind speech separation based on AuxIVA and incomplete demixing transform}, 
  year={2016},
  volume={},
  number={},
  pages={1-5},
  doi={10.1109/IWAENC.2016.7602921}
}

@ARTICLE{Hyvarinen2000ISA,
  author={Hyvärinen, Aapo and Hoyer, Patrik},
  journal={Neural Computation}, 
  title={Emergence of phase- and shift-invariant features by decomposition of natural images into independent feature subspaces}, 
  year={2000},
  volume={12},
  number={7},
  pages={1705-1720},
  doi={10.1162/089976600300015312}
}

@book{benveniste2012adaptive,
  title={Adaptive algorithms and stochastic approximations},
  author={Benveniste, Albert and M{\'e}tivier, Michel and Priouret, Pierre},
  volume={},
  year={1990},
  publisher={Springer Berlin, Heidelberg},
  doi = {https://doi.org/10.1007/978-3-642-75894-2}
}

@book{anderson2013thesis,
  title={Independent vector analysis: Theory, algorithms, and applications},
  author={Anderson, Matthew},
  year={2013},
  publisher={University of Maryland, Baltimore County}
}

@INPROCEEDINGS{IkeshitaOverIVA,
  author={Ikeshita, Rintaro and Nakatani, Tomohiro and Araki, Shoko},
  booktitle={ICASSP 2020 - 2020 IEEE International Conference on Acoustics, Speech and Signal Processing (ICASSP)}, 
  title={Overdetermined independent vector analysis}, 
  year={2020},
  volume={},
  number={},
  pages={591-595},
  doi={10.1109/ICASSP40776.2020.9053790}
}

@article{GABRIELSON2020101948,
title = {Joint-IVA for identification of discriminating features in EEG: Application to a driving study},
journal = {Biomedical Signal Processing and Control},
volume = {61},
pages = {101948},
year = {2020},
issn = {1746-8094},
doi = {https://doi.org/10.1016/j.bspc.2020.101948},
url = {https://www.sciencedirect.com/science/article/pii/S174680942030104X},
author = {Ben Gabrielson and M.A.B.S. Akhonda and Suchita Bhinge and Justin Brooks and Qunfang Long and Tülay Adali},
}

@ARTICLE{Li2022comIVA,
  author={Li, Mingchun and Chang, Zhengwei and Zhang, Linghao and Xu, Houdong and Luo, Zhongqiang and Guo, Ruiming},
  journal={IEEE Access}, 
  title={Blind separation for wireless communication convolutive mixtures based on denoising IVA}, 
  year={2022},
  volume={10},
  number={},
  pages={113756-113766},
  doi={10.1109/ACCESS.2022.3218633}
}

@INPROCEEDINGS{Amor2021,
  author={Amor, N. and Čmejla, J. and Kautský, V. and Koldovský, Z. and Kounovský, T.},
  booktitle={ICASSP 2021 - 2021 IEEE International Conference on Acoustics, Speech and Signal Processing (ICASSP)}, 
  title={Blind extraction of moving sources via independent component and vector analysis: Examples}, 
  year={2021},
  volume={},
  number={},
  pages={3725-3729},
  doi={10.1109/ICASSP39728.2021.9413422}
}

@Article{Luo2022com,
AUTHOR = {Luo, Zhongqiang and Guo, Ruiming and Li, Chengjie},
TITLE = {Independent vector analysis for blind deconvolving of digital modulated communication signals},
JOURNAL = {Electronics},
VOLUME = {11},
YEAR = {2022},
NUMBER = {9},
ARTICLE-NUMBER = {1460},
URL = {https://www.mdpi.com/2079-9292/11/9/1460},
ISSN = {2079-9292},
DOI = {10.3390/electronics11091460}
}

@ARTICLE{Luo2022mimo,
  author={Luo, Zhongqiang and Li, Mingchun and Li, Chengjie},
  journal={China Communications}, 
  title={Independent vector analysis based blind interference reduction and signal recovery for MIMO IoT green communications}, 
  year={2022},
  volume={19},
  number={7},
  pages={79-88},
  doi={10.23919/JCC.2022.07.007}
}

@article{LEE2008fmri,
title = {Independent vector analysis (IVA): Multivariate approach for fMRI group study},
journal = {NeuroImage},
volume = {40},
number = {1},
pages = {86-109},
year = {2008},
issn = {1053-8119},
doi = {https://doi.org/10.1016/j.neuroimage.2007.11.019},
author = {Jong-Hwan Lee and Te-Won Lee and Ferenc A. Jolesz and Seung-Schik Yoo},
}

@Inbook{Lee2007speech,
author="Lee, Intae
and Kim, Taesu
and Lee, Te-Won",
editor="Makino, Shoji
and Sawada, Hiroshi
and Lee, Te-Won",
title="Independent vector analysis for convolutive blind speech separation",
bookTitle="Blind Speech Separation",
year="2007",
publisher="Springer Netherlands",
address="Dordrecht",
pages="169-192",
isbn="978-1-4020-6479-1",
doi="10.1007/978-1-4020-6479-1_6",
url="https://doi.org/10.1007/978-1-4020-6479-1_6"
}

@ARTICLE{Adali2018,
  author={Adali, Tülay and Akhonda, M. A. B. S. and Calhoun, Vince D.},
  journal={IEEE Sensors Letters}, 
  title={ICA and IVA for data fusion: An overview and a new approach based on disjoint subspaces}, 
  year={2019},
  volume={3},
  number={1},
  pages={1-4},
  doi={10.1109/LSENS.2018.2884775}
}

@INPROCEEDINGS{Brendel2020spat,
  author={Brendel, Andreas and Haubner, Thomas and Kellermann, Walter},
  booktitle={ICASSP 2020 - 2020 IEEE International Conference on Acoustics, Speech and Signal Processing (ICASSP)}, 
  title={Spatially guided independent vector analysis}, 
  year={2020},
  volume={},
  number={},
  pages={596-600},
  doi={10.1109/ICASSP40776.2020.9052905}
}

@book{HyvarinenICAbook,
title = "Independent Component Analysis",
author = "Aapo Hyv{\"a}rinen and J. Karhunen and E. Oja",
year = "2001",
doi = "10.1002/0471221317",
language = "English",
isbn = "9780471405405",
publisher = "Wiley",
address = "United Kingdom",
}

@inproceedings{Hastie2002ProDensICA,
author = {Hastie, Trevor and Tibshirani, Rob},
title = {Independent components analysis through product density estimation},
year = {2002},
publisher = {MIT Press},
address = {Cambridge, MA, USA},
booktitle = {Proceedings of the 16th International Conference on Neural Information Processing Systems},
pages = {665–672},
numpages = {8},
series = {NIPS'02}
}

@article{Samworth2012logconcICA,
author = {Richard J. Samworth and Ming Yuan},
title = {{Independent component analysis via nonparametric maximum likelihood estimation}},
volume = {40},
journal = {The Annals of Statistics},
number = {6},
publisher = {Institute of Mathematical Statistics},
pages = {2973 - 3002},
year = {2012},
doi = {10.1214/12-AOS1060},
URL = {https://doi.org/10.1214/12-AOS1060}
}

@inproceedings{karvanen2000maximum,
  title={Maximum likelihood estimation of ICA model for wide class of source distributions},
  author={Karvanen, Juha and Eriksson, Jan and Koivunen, Visa},
  booktitle={Neural Networks for Signal Processing X. Proceedings of the 2000 IEEE Signal Processing Society Workshop (Cat. No. 00TH8501)},
  volume={1},
  pages={445-454},
  year={2000},
doi ={10.1109/NNSP.2000.889437},
  organization={IEEE}
}

@Article{Guo2023review,
AUTHOR = {Guo, Ruiming and Luo, Zhongqiang and Li, Mingchun},
TITLE = {A survey of optimization methods for independent vector analysis in audio source separation},
JOURNAL = {Sensors},
VOLUME = {23},
YEAR = {2023},
NUMBER = {1},
ARTICLE-NUMBER = {493},
URL = {https://www.mdpi.com/1424-8220/23/1/493},
PubMedID = {36617090},
ISSN = {1424-8220},
DOI = {10.3390/s23010493}
}

@ARTICLE{Li2007,
  author={Li, Xi-Lin and Zhang, Xian-Da},
  journal={IEEE Transactions on Signal Processing}, 
  title={Nonorthogonal joint diagonalization free of degenerate solution}, 
  year={2007},
  volume={55},
  number={5},
  pages={1803-1814},
  doi={10.1109/TSP.2006.889983}
}

@ARTICLE{Lahat2019JISAidentifiability,
  author={Lahat, Dana and Jutten, Christian},
  journal={IEEE Transactions on Signal Processing}, 
  title={Joint independent subspace analysis: uniqueness and identifiability}, 
  year={2019},
  volume={67},
  number={3},
  pages={684-699},
  doi={10.1109/TSP.2018.2880714}
}

@INPROCEEDINGS{Lahat2014jisa,
  author={Lahat, Dana and Jutten, Christian},
  booktitle={2014 22nd European Signal Processing Conference (EUSIPCO)}, 
  title={Joint blind source separation of multidimensional components: Model and algorithm}, 
  year={2014},
  volume={},
  number={},
  pages={1417-1421},
  doi={}
}

@article{Kettering1971,
 ISSN = {00063444, 14643510},
 URL = {http://www.jstor.org/stable/2334380},
 author = {J. R. Kettenring},
 journal = {Biometrika},
 number = {3},
 pages = {433-451},
 publisher = {[Oxford University Press, Biometrika Trust]},
 title = {Canonical analysis of several sets of variables},
 urldate = {2025-03-14},
 volume = {58},
doi={10.2307/2334380},
 year = {1971}
}

@article{Calhoun2006jica,
author = {Calhoun, V.D. and Adali, T. and Giuliani, N.R. and Pekar, J.J. and Kiehl, K.A. and Pearlson, G.D.},
title = {Method for multimodal analysis of independent source differences in schizophrenia: Combining gray matter structural and auditory oddball functional data},
journal = {Human Brain Mapping},
volume = {27},
number = {1},
pages = {47-62},
doi = {https://doi.org/10.1002/hbm.20166},
url = {https://onlinelibrary.wiley.com/doi/abs/10.1002/hbm.20166},
eprint = {https://onlinelibrary.wiley.com/doi/pdf/10.1002/hbm.20166},
year = {2006}
}

@article{Calhoun2001,
author = {Calhoun, V.D. and Adali, T. and Pearlson, G.D. and Pekar, J.J.},
title = {A method for making group inferences from functional MRI data using independent component analysis},
journal = {Human Brain Mapping},
volume = {14},
number = {3},
pages = {140-151},
doi = {https://doi.org/10.1002/hbm.1048},
url = {https://onlinelibrary.wiley.com/doi/abs/10.1002/hbm.1048},
year = {2001}
}

@article{BECKMANN2005294,
title = {Tensorial extensions of independent component analysis for multisubject FMRI analysis},
journal = {NeuroImage},
volume = {25},
number = {1},
pages = {294-311},
year = {2005},
issn = {1053-8119},
doi = {https://doi.org/10.1016/j.neuroimage.2004.10.043},
url = {https://www.sciencedirect.com/science/article/pii/S1053811904006378},
author = {C.F. Beckmann and S.M. Smith}
}

@incollection{ONTON200699,
title = {Information-based modeling of event-related brain dynamics},
editor = {Christa Neuper and Wolfgang Klimesch},
series = {Progress in Brain Research},
publisher = {Elsevier},
volume = {159},
pages = {99-120},
year = {2006},
booktitle = {Event-Related Dynamics of Brain Oscillations},
issn = {0079-6123},
doi = {https://doi.org/10.1016/S0079-6123(06)59007-7},
url = {https://www.sciencedirect.com/science/article/pii/S0079612306590077},
author = {Julie Onton and Scott Makeig}
}

@article{Nielsin2002,
author = {Nielsen, Allan},
year = {2002},
month = {04},
pages = {293 - 305},
title = {Multiset canonical correlations analysis and multispectral, truly multitemporal remote sensing data},
volume = {11},
journal = {IEEE Transactions on Image Processing},
doi = {10.1109/83.988962}
}

@ARTICLE{Lahat2016JISA,
  author={Lahat, Dana and Jutten, Christian},
  journal={IEEE Transactions on Signal Processing}, 
  title={Joint independent subspace analysis using second-order statistics}, 
  year={2016},
  volume={64},
  number={18},
  pages={4891-4904},
  doi={10.1109/TSP.2016.2526960}}

@inproceedings{Theis2006,
  title={Towards a general independent subspace analysis},
  author={Fabian J Theis},
  booktitle={Neural Information Processing Systems},
  year={2006},
  url={https://api.semanticscholar.org/CorpusID:15361445}
}

@article{CALHOUN2009,
title = {A review of group ICA for fMRI data and ICA for joint inference of imaging, genetic, and ERP data},
journal = {NeuroImage},
volume = {45},
number = {1, Supplement 1},
pages = {S163-S172},
year = {2009},
note = {Mathematics in Brain Imaging},
issn = {1053-8119},
doi = {https://doi.org/10.1016/j.neuroimage.2008.10.057},
url = {https://www.sciencedirect.com/science/article/pii/S1053811908012032},
author = {Vince D. Calhoun and Jingyu Liu and Tülay Adalı}
}

@article{Esposito2005,
author = {Esposito, Fabrizio and Scarabino, Tommaso and Hyvarinen, Aapo and Himberg, Johan and Formisano, Elia and Comani, Silvia and Tedeschi, Gioacchino and Goebel, Rainer and Seifritz, Erich and Salle, Francesco},
year = {2005},
month = {04},
pages = {193-205},
title = {Independent component analysis of fmri group studies by self-organizing clustering},
volume = {25},
journal = {NeuroImage},
doi = {10.1016/j.neuroimage.2004.10.042}
}

@INPROCEEDINGS{Ying2008,
  author={Guo, Ying},
  booktitle={2008 International Conference on BioMedical Engineering and Informatics}, 
  title={Group independent component analysis of multi-subject fMRI data: Connections and distinctions between two methods}, 
  year={2008},
  volume={2},
  number={},
  pages={748-752},
  doi={10.1109/BMEI.2008.191}
}

@Article{Guo2023,
AUTHOR = {Guo, Ruiming and Luo, Zhongqiang and Wang, Ling and Feng, Li},
TITLE = {Efficient overdetermined independent vector analysis based on iterative projection with adjustment},
JOURNAL = {Electronics},
VOLUME = {12},
YEAR = {2023},
NUMBER = {14},
ARTICLE-NUMBER = {3200},
URL = {https://www.mdpi.com/2079-9292/12/14/3200},
ISSN = {2079-9292},
DOI = {10.3390/electronics12143200}
}

@INPROCEEDINGS{Goto2022,
  author={Goto, Kana and Ueda, Tetsuya and Li, Li and Yamada, Takeshi and Makino, Shoji},
  booktitle={2022 Asia-Pacific Signal and Information Processing Association Annual Summit and Conference (APSIPA ASC)}, 
  title={Accelerating online algorithm using geometrically constrained independent vector analysis with iterative source steering}, 
  year={2022},
  volume={},
  number={},
  pages={754-759},
  doi={10.23919/APSIPAASC55919.2022.9980301}
}

@INPROCEEDINGS{Ono2012,
  author={Ono, Nobutaka},
  booktitle={Proceedings of The 2012 Asia Pacific Signal and Information Processing Association Annual Summit and Conference}, 
  title={Auxiliary-function-based independent vector analysis with power of vector-norm type weighting functions}, 
  year={2012},
  volume={},
  number={},
  pages={1-4},
  doi={}
}

@article{Bingham2000,
title = "A fast fixed-point algorithm for independent component analysis of complex valued signals",
author = "E. Bingham and Aapo Hyv{\"a}rinen",
year = "2000",
language = "English",
volume = "10",
pages = "1-8",
journal = "International Journal of Neural Systems",
issn = "1793-6462",
publisher = "World Scientific",
number = "1",
doi = "10.1142/S0129065700000028",
}

@INPROCEEDINGS{Calhoun2002,
  author={Calhoun, V. and Adali, T.},
  booktitle={Proceedings of the 12th IEEE Workshop on Neural Networks for Signal Processing}, 
  title={Complex infomax: convergence and approximation of infomax with complex nonlinearities}, 
  year={2002},
  volume={},
  number={},
  pages={307-316},
  doi={10.1109/NNSP.2002.1030042}
}

@article{Koldovsky2021,
  author={Koldovský, Zbyněk and Kautský, Václav and Tichavský, Petr and Čmejla, Jaroslav and Málek, Jiří},
  journal={IEEE Transactions on Signal Processing}, 
  title={Dynamic independent component/vector analysis: Time-variant linear mixtures separable by time-invariant beamformers}, 
  year={2021},
  volume={69},
  number={},
  pages={2158-2173},
  doi={10.1109/TSP.2021.3068626}
}

@article{Kullback1951,
 ISSN = {00034851},
 URL = {http://www.jstor.org/stable/2236703},
 author = {S. Kullback and R. A. Leibler},
 journal = {The Annals of Mathematical Statistics},
 number = {1},
 pages = {79-86},
 publisher = {Institute of Mathematical Statistics},
 title = {On information and sufficiency},
 volume = {22},
 year = {1951},
doi = {10.1214/aoms/1177729694}
}

@ARTICLE{Vincent2006,
  author={Vincent, E. and Gribonval, R. and Fevotte, C.},
  journal={IEEE Transactions on Audio, Speech, and Language Processing}, 
  title={Performance measurement in blind audio source separation}, 
  year={2006},
  volume={14},
  number={4},
  pages={1462-1469},
  doi={10.1109/TSA.2005.858005}
}

@inproceedings{Amari1995,
 author = {Amari, Shun-ichi and Cichocki, Andrzej and Yang, Howard},
 booktitle = {Advances in Neural Information Processing Systems},
 editor = {D. Touretzky and M.C. Mozer and M. Hasselmo},
 pages = {757-763},
 title = {A new learning algorithm for blind signal separation},
 url = {https://proceedings.neurips.cc/paper_files/paper/1995/file/e19347e1c3ca0c0b97de5fb3b690855a-Paper.pdf},
 volume = {8},
 year = {1996}
}

@INPROCEEDINGS{Boukouvalas2015,
  author={Boukouvalas, Zois and Fu, Geng-Shen and Adalı, Tülay},
  booktitle={2015 49th Annual Conference on Information Sciences and Systems (CISS)}, 
  title={An efficient multivariate generalized Gaussian distribution estimator: Application to IVA}, 
  year={2015},
  volume={},
  number={},
  pages={1-4},
  doi={10.1109/CISS.2015.7086828}
}

@INPROCEEDINGS{Damasceno2021,
  author={Damasceno, Lucas P. and Cavalcante, Charles C. and Adalı, Tülay and Boukouvalas, Zois},
  booktitle={ICASSP 2021 - 2021 IEEE International Conference on Acoustics, Speech and Signal Processing (ICASSP)}, 
  title={Independent vector analysis using semi-parametric density estimation via multivariate entropy maximization}, 
  year={2021},
  volume={},
  number={},
  pages={3715-3719},
  doi={10.1109/ICASSP39728.2021.9414839}
}

@inproceedings{Scheibler2019,
author={Scheibler, Robin and Ono, Nobutaka},
  booktitle={2019 IEEE Workshop on Applications of Signal Processing to Audio and Acoustics (WASPAA)}, 
  title={Independent vector analysis with more microphones than sources}, 
  year={2019},
  volume={},
  number={},
  pages={185-189},
  doi={10.1109/WASPAA.2019.8937080}}

@InProceedings{Taesu2006IVA,
author="Kim, Taesu
and Eltoft, Torbj{\o}rn
and Lee, Te-Won",
editor="Rosca, Justinian
and Erdogmus, Deniz
and Pr{\'i}ncipe, Jos{\'e} C.
and Haykin, Simon",
title="Independent vector analysis: An extension of ICA to multivariate components",
booktitle="Independent Component Analysis and Blind Signal Separation",
year="2006",
publisher="Springer Berlin Heidelberg",
address="Berlin, Heidelberg",
pages="165-172",
isbn="978-3-540-32631-1",
doi = "10.1007/11679363_21"
}

@INPROCEEDINGS{Andreas2021,
  author={Brendel, Andreas and Kellermann, Walter},
  booktitle={2021 IEEE Workshop on Applications of Signal Processing to Audio and Acoustics (WASPAA)}, 
  title={Fasteriva: Update rules for independent vector analysis based on negentropy and the majorize-minimize principle}, 
  year={2021},
  volume={},
  number={},
  pages={131-135},
  doi={10.1109/WASPAA52581.2021.9632790}
}

@InProceedings{Anderson2010trick,
author="Anderson, Matthew
and Li, Xi-Lin
and Adali, T{\"u}lay",
editor="Vigneron, Vincent
and Zarzoso, Vicente
and Moreau, Eric
and Gribonval, R{\'e}mi
and Vincent, Emmanuel",
title="Nonorthogonal independent vector analysis using multivariate Gaussian model",
booktitle="Latent Variable Analysis and Signal Separation",
year="2010",
publisher="Springer Berlin Heidelberg",
address="Berlin, Heidelberg",
pages="354-361",
isbn="978-3-642-15995-4",
doi = "10.1007/978-3-642-15995-4_44"
}

@inproceedings{Hyvarinen1997negappox,
author = {Hyv\"{a}rinen, Aapo},
title = {New approximations of differential entropy for independent component analysis and projection pursuit},
year = {1997},
publisher = {MIT Press},
address = {Cambridge, MA, USA},
booktitle = {Proceedings of the 11th International Conference on Neural Information Processing Systems},
pages = {273–279},
numpages = {7},
location = {Denver, CO},
series = {NIPS'97}
}

@article{Jones1987,
author = {Jones, M.C. and Sibson, Robin},
title = {What is projection pursuit?},
journal = {Journal of the Royal Statistical Society: Series A},
volume = {150},
number = {1},
pages = {1-18},
doi = {10.2307/2981662},
url = {https://rss.onlinelibrary.wiley.com/doi/abs/10.2307/2981662},
eprint = {https://rss.onlinelibrary.wiley.com/doi/pdf/10.2307/2981662},
year = {1987}
}

@INPROCEEDINGS{Li2020geom,
  author={Li, Li and Koishida, Kazuhito},
  booktitle={ICASSP 2020 - 2020 IEEE International Conference on Acoustics, Speech and Signal Processing (ICASSP)}, 
  title={Geometrically constrained independent vector analysis for directional speech enhancement}, 
  year={2020},
  volume={},
  number={},
  pages={846-850},
  doi={10.1109/ICASSP40776.2020.9053649}}

@article{Scheibler2021,
author = {Scheibler, Robin},
year = {2021},
month = {04},
pages = {2509-2524},
title = {Independent vector analysis via log-quadratically penalized quadratic minimization},
volume = {69},
journal = {IEEE Transactions on Signal Processing},
doi = {10.1109/TSP.2021.3072228}
}

@article{Gomez1998,
author = {Gómez, E. and Gómez-Villegas, Miguel and Marin, J.},
year = {1998},
month = {01},
pages = {589-600},
title = {A multivariate generalization of the power exponential family of distributions},
volume = {27},
journal = {Communications in Statistics - Theory and Methods},
doi = {10.1080/03610929808832115}
}

@INPROCEEDINGS{Via2011,
  author={Vía, Javier and Anderson, Matthew and Li, Xi-Lin and Adalı, Tülay},
  booktitle={2011 IEEE International Conference on Acoustics, Speech and Signal Processing (ICASSP)}, 
  title={Joint blind source separation from second-order statistics: Necessary and sufficient identifiability conditions}, 
  year={2011},
  volume={},
  number={},
  pages={2520-2523},
  doi={10.1109/ICASSP.2011.5946997}
}

@INPROCEEDINGS{Qian2014,
  author={Qian, Guobing and Li, Liping and Liao, Hongshu},
  booktitle={2014 IEEE International Conference on Communiction Problem-solving}, 
  title={Stable analysis of fast independent vector analysis algorithm}, 
  year={2014},
  volume={},
  number={},
  pages={485-488},
  doi={10.1109/ICCPS.2014.7062328}
}

@inproceedings{Ono2010,
author="Ono, Nobutaka
and Miyabe, Shigeki",
editor="Vigneron, Vincent
and Zarzoso, Vicente
and Moreau, Eric
and Gribonval, R{\'e}mi
and Vincent, Emmanuel",
title="Auxiliary-function-based independent component analysis for super-Gaussian sources",
booktitle="Latent Variable Analysis and Signal Separation",
year="2010",
publisher="Springer Berlin Heidelberg",
pages="165-172",
doi = {10.1007/978-3-642-15995-4_21}
}

@inproceedings{Kang2021,
  author={Kang, Fang and Yang, Feiran and Yang, Jun},
  booktitle={2021 IEEE Spoken Language Technology Workshop (SLT)}, 
  title={Real-time independent vector analysis with a deep-learning-based source model}, 
  year={2021},
  volume={},
  number={},
  pages={665-669},
  doi={10.1109/SLT48900.2021.9383599}
}

@article{Li2020IndependentVA,
  title={Independent Vector Analysis with Deep Neural Network Source Priors},
  author={Xi-Lin Li},
  journal={arXiv: Audio and Speech Processing},
  year={2020},
  url={https://api.semanticscholar.org/CorpusID:222175849}
}

@article{Correa2008,
  author={Correa, Nicolle M. and Li, Yi-Ou and Adali, TÜlay and Calhoun, Vince D.},
  journal={IEEE Journal of Selected Topics in Signal Processing}, 
  title={Canonical correlation analysis for feature-based fusion of biomedical imaging modalities and its application to detection of associative networks in schizophrenia}, 
  year={2008},
  volume={2},
  number={6},
  pages={998-1007},
  doi={10.1109/JSTSP.2008.2008265}
}

@article{Adali2015,
  author={Adali, Tülay and Levin-Schwartz, Yuri and Calhoun, Vince D.},
  journal={Proceedings of the IEEE}, 
  title={Multimodal data fusion using source separation: Two effective models based on ICA and IVA and their properties}, 
  year={2015},
  volume={103},
  number={9},
  pages={1478-1493},
  doi={10.1109/JPROC.2015.2461624}
}

@article{Pearlmutter1996,
  title={Maximum likelihood blind source separation: A context-sensitive generalization of ICA},
  author={Pearlmutter, Barak and Parra, Lucas},
  journal={Advances in Neural Information Processing Systems},
  volume={9},
  pages={613-619},
  year={1996}
}

@article{Cardoso1997,
  author={Cardoso, J.-F.},
  journal={IEEE Signal Processing Letters}, 
  title={Infomax and maximum likelihood for blind source separation}, 
  year={1997},
  volume={4},
  number={4},
  pages={112-114},
  doi={10.1109/97.566704}
}

@article{Bell1995,
author = {Bell, Anthony and Sejnowski, Terrence},
year = {1995},
month = {12},
pages = {1129-59},
title = {An information-maximization approach to blind separation and blind deconvolution},
volume = {7},
journal = {Neural computation},
doi = {10.1162/neco.1995.7.6.1129}
}

@article{Nadal1995,
author = {Nadal, Jean-Pierre and Parga, Nestor},
year = {1995},
month = {03},
pages = {},
title = {Non linear neurons in the low noise limit: A factorial code maximizes information transfer},
volume = {5},
journal = {Network Computation in Neural Systems},
doi = {10.1088/0954-898X/5/4/008}
}

@article{LIANG2014175,
title = {Independent vector analysis with a generalized multivariate Gaussian source prior for frequency domain blind source separation},
journal = {Signal Processing},
volume = {105},
pages = {175-184},
year = {2014},
issn = {0165-1684},
doi = {10.1016/j.sigpro.2014.05.022},
url = {https://www.sciencedirect.com/science/article/pii/S0165168414002400},
author = {Yanfeng Liang and Jack Harris and Syed Mohsen Naqvi and Gaojie Chen and Jonathon A. Chambers}
}

@inproceedings{Rafique2016,
author={Rafique, Waqas and Erateb, Suleiman and Naqvi, Syed Mohsen and Dlay, Satnam S. and Chambers, Jonathon A.},
  booktitle={2016 24th European Signal Processing Conference (EUSIPCO)}, 
  title={Independent vector analysis for source separation using an energy driven mixed student's T and super Gaussian source prior}, 
  year={2016},
  volume={},
  number={},
  pages={858-862},
  doi={10.1109/EUSIPCO.2016.7760370}}

@article{Liang2013,
author = {Liang, Yanfeng and Chen, Gaojie and Naqvi, Syed and Chambers, Jonathon},
year = {2013},
month = {08},
pages = {},
title = {Independent vector analysis with multivariate student's t-distribution source prior for speech separation},
volume = {49},
journal = {Electronics Letters},
doi = {10.1049/el.2013.1999}
}

@article{Arslan2010,
author = {Arslan, Olcay},
year = {2010},
month = {12},
pages = {865-887},
title = {An alternative multivariate skew Laplace distribution: Properties and estimation},
volume = {51},
journal = {Statistical Papers},
doi = {10.1007/s00362-008-0183-7}
}

@INPROCEEDINGS{Anderson2013kotz,
  author={Anderson, Matthew and Fu, Geng-Shen and Phlypo, Ronald and Adali, Tülay},
  booktitle={2013 IEEE International Conference on Acoustics, Speech and Signal Processing}, 
  title={Independent vector analysis, the Kotz distribution, and performance bounds}, 
  year={2013},
  volume={},
  number={},
  pages={3243-3247},
  doi={10.1109/ICASSP.2013.6638257}
}

@ARTICLE{Anderson2012,
  author={Anderson, Matthew and Adali, Tülay and Li, Xi-Lin},
  journal={IEEE Transactions on Signal Processing}, 
  title={Joint blind source separation with multivariate Gaussian model: Algorithms and performance analysis}, 
  year={2012},
  volume={60},
  number={4},
  pages={1672-1683},
  doi={10.1109/TSP.2011.2181836}
}

@inproceedings{Ono2011,
author = {Ono, Nobutaka},
year = {2011},
pages = {189-192},
title = {Stable and fast update rules for independent vector analysis based on auxiliary function technique},
booktitle={2011 IEEE Workshop on Applications of Signal Processing to Audio and Acoustics (WASPAA)}
}

@article{anderson2013,
author = {Anderson, Matthew and Fu, Geng-Shen and Phlypo, Ronald and Adali, Tülay},
year = {2013},
pages={4399-4410},
title = {Independent vector analysis: Identification conditions and performance bounds},
volume = {62},
number={17},
journal = {IEEE Transactions on Signal Processing},
doi = {10.1109/TSP.2014.2333554}
}

@INPROCEEDINGS{taesu2006,
  author={Kim, Taesu and Lee, Intae and Lee, Te-Won},
  booktitle={2006 Fortieth Asilomar Conference on Signals, Systems and Computers}, 
  title={Independent vector analysis: Definition and algorithms}, 
  year={2006},
  volume={},
  number={},
  pages={1393-1396},
doi = {10.1109/ACSSC.2006.354986}
}

@article{lee2007,
author = {Lee, Intae and Kim, Taesu and Lee, Te-Won},
year = {2007},
pages = {1859-1871},
title = {Fast fixed-point independent vector analysis algorithms for convolutive blind source separation},
volume = {87},
journal = {Signal Processing},
doi = {10.1016/j.sigpro.2007.01.010}
}

@article{HYVARINEN2000411,
title = {Independent component analysis: algorithms and applications},
journal = {Neural Networks},
volume = {13},
number = {4},
pages = {411-430},
year = {2000},
issn = {0893-6080},
author = {A. Hyvärinen and E. Oja},
doi = {10.1016/S0893-6080(00)00026-5}
}

@article{Mowakeaa2020,
author = {Mowakeaa, Rami and Boukouvalas, Zois and Long, Qunfang and Adali, Tülay},
year = {2020},
title = {{IVA} using complex multivariate {GGD}: application to f{MRI} analysis},
volume = {31},
journal = {Multidimensional Systems and Signal Processing},
pages={725-744},
publisher={Springer}
}

@article{nordhausen2019overview,
  title={An overview of properties and extensions of FOBI},
  author={Nordhausen, Klaus and Virta, Joni},
  journal={Knowledge-Based Systems},
  volume={173},
  pages={113-116},
  year={2019},
  publisher={Elsevier},
  doi = {10.1016/j.knosys.2019.02.026}
}

@article{miettinen2015,
author = "Miettinen, Jari and Taskinen, Sara and Nordhausen, Klaus and Oja, Hannu",
journal = "Statistical Science",
fjournal = "Statist. Sci.",
number = "3",
pages = "372-390",
publisher = "The Institute of Mathematical Statistics",
title = "Fourth moments and independent component analysis",
volume = "30",
year = "2015",
doi = "10.1214/15-STS520"
}

@article{OjaSirkiaEriksson2006, 
	title={Scatter matrices and independent component analysis}, 
	volume={35}, 
	journal={Austrian Journal of Statistics}, 
	author={Oja, Hannu and Sirkiä, Seija and Eriksson, Jan}, 
	year={2006},
	pages={175–189},
doi = {10.17713/ajs.v35i2&3.364}
}

@inproceedings{Cardoso1989,
author = {Cardoso, J.-F.},
title = {Source separation using higher order moments},
booktitle = {Proceedings of the IEEE International Conference on Acoustics, Speech and Signal Processing},
pages = {2109-2112},
year = {1989},
doi = {10.1109/ICASSP.1989.266878}
}

@article{NordhausenOja2018,
author = {Nordhausen, K. and Oja, H.},
title = {Independent component analysis: A statistical perspective},
journal = {WIREs: Computational Statistics},
volume = {10},
pages = {e1440},
year = {2018},
doi = {10.1002/wics.1440}
}

@Manual{R,
    title = {\textsf{R}: A Language and Environment for Statistical Computing},
    author = {{\textsf{R} Core Team}},
    organization = {\textsf{R} Foundation for Statistical Computing},
    address = {Vienna, Austria},
    year = {2021},
    url = {https://www.R-project.org/},
  }

@book{ComonJutten2010,
    title={Handbook of Blind Source Separation: {I}ndependent Component Analysis and Applications},
    author={Comon, P. and Jutten, C.},
    year={2010},
    publisher={Academic Press},
    address = {Oxford},
    doi={10.1016/C2009-0-19334-0}
}

@article{virta2017independent,
	title={Independent component analysis for tensor-valued data},
	author={Virta, Joni and Li, Bing and Nordhausen, Klaus and Oja, Hannu},
	journal={Journal of Multivariate Analysis},
	volume={162},
	pages={172-192},
	year={2017},
doi = {10.1016/j.jmva.2017.09.008}
}

@article{Hotelling1936,
    author = {Hotelling, Harold},
    title = {Relations between two sets of variates},
    journal = {Biometrika},
    volume = {28},
    pages = {321-377},
    year = {1936},
    doi = {10.1093/biomet/28.3-4.321},
}

@article{Hyvarinen:1999,
  author = {Hyv\"{a}rinen, A.},
  title = {Fast and robust fixed-point algorithms for independent component analysis},
  journal = {IEEE Transactions on Neural Networks},
  year = 1999,
  volume = 10,
  doi = {10.1109/72.761722},
  pages = {626-634}}

@article{HyvarinenOja:1997,
  author = {Hyv{\"a}rinen, A. and Oja, E.},
  title = {A Fast Fixed-Point Algorithm for Independent Component Analyis},
  journal = {Neural Computation},
  year = {1997},
  volume = {9},
  pages = {1483-1492},
  doi={ 10.1162/neco.1997.9.7.1483}
  }

@ARTICLE{CardosoSouloumiac1993,
   author = {Cardoso, J.-F.  and Souloumiac,  A. },
   title = {Blind beamforming for non-{G}aussian signals},
   journal = {IEE Proceedings F},  
   volume = {140},
   year = {1993},
   pages = {362-370}
}

@article{Miettinenetal:2017,
    author={Miettinen, J. and Nordhausen, K. and Oja, H. and Taskinen, S. and Virta, J.},
    journal={Signal Processing},
    title={The squared symmetric {FastICA} estimator},
    year={2017},
    volume={131},
    pages={402-411},
    doi={10.1016/j.sigpro.2016.08.028}
    }

@ARTICLE{Ollila:2010,
  author={Ollila, Esa},
  journal={IEEE Transactions on Signal Processing}, 
  title={The deflation-based {FastICA} estimator: Statistical Analysis Revisited}, 
  year={2010},
  volume={58},
  number={3},
  pages={1527-1541},
  doi={10.1109/TSP.2009.2036072}}

@INPROCEEDINGS{Nordhausenetal:2011,
  author={Nordhausen, Klaus and Ilmonen, Pauliina and Mandal, Abhijit and Oja, Hannu and Ollila, Esa},
  booktitle={2011 19th European Signal Processing Conference}, 
  title={Deflation-based {FastICA} reloaded}, 
  year={2011},
  volume={},
  number={},
  pages={1854-1858},
  doi={}}

@article{Comon1994,
title = {Independent component analysis, A new concept?},
journal = {Signal Processing},
volume = {36},
number = {3},
pages = {287-314},
year = {1994},
doi = {10.1016/0165-1684(94)90029-9},
author = {Pierre Comon},
}

@article{NordhausenRuizGazen2022,
title = {On the usage of joint diagonalization in multivariate statistics},
journal = {Journal of Multivariate Analysis},
volume = {188},
pages = {104844},
year = {2022},
doi = {10.1016/j.jmva.2021.104844},
author = {Klaus Nordhausen and Anne Ruiz-Gazen},
}

@article{VirtaLiNordhausenOja2020,
title = {Independent component analysis for multivariate functional data},
journal = {Journal of Multivariate Analysis},
volume = {176},
pages = {104568},
year = {2020},
doi = {10.1016/j.jmva.2019.104568},
author = {Joni Virta and Bing Li and Klaus Nordhausen and Hannu Oja},
}

@article{MiettinenNordhausenOjaTaskinen2014,
title = "Deflation-based FastICA with adaptive choices of nonlinearities",
author = "Jari Miettinen and Klaus Nordhausen and Hannu Oja and Sara Taskinen",
year = "2014",
volume = "62",
pages = "5716-5724",
journal = "IEEE Transactions on Signal Processing",
number = "21",
doi = "10.1109/TSP.2014.2356442"
}

\end{document}